\documentclass[journal]{IEEEtran}

\usepackage{booktabs}
\usepackage{threeparttable}
\usepackage{amsmath,graphicx,cite,amssymb}
\usepackage{amsfonts,multirow,bm,array,setspace,stfloats}

\usepackage{graphicx,float,cite,amssymb}
\usepackage{graphics}
 \DeclareGraphicsExtensions{.pdf,.jpeg,.png,.jpg}   
\usepackage{multirow,bm,bbm,array,setspace,dsfont}
\usepackage{textcomp}
\usepackage{psfrag}
\usepackage{color}
\usepackage{pstricks,enumerate}
\usepackage{bbm}
\usepackage{subfigure}

\usepackage{algorithm,algpseudocode}
\makeatletter
\newcommand{\distas}[1]{\mathbin{\overset{#1}{\kern\z@\sim}}}%
\newsavebox{\mybox}\newsavebox{\mysim}
\newcommand{\distras}[1]{%
  \savebox{\mybox}{\hbox{\kern1pt$\scriptstyle#1$\kern1pt}}%
  \savebox{\mysim}{\hbox{$\sim$}}%
  \mathbin{\overset{#1}{\kern\z@\resizebox{\wd\mybox}{\ht\mysim}{$\sim$}}}%
}
\makeatother
\ifCLASSINFOpdf
\else
\fi

\newtheorem{proposition}{Proposition}

\newtheorem{lemma}{Lemma}
\newtheorem{theorem}{Theorem}
\newtheorem{corollary}{Corollary}
\newtheorem{example}{Example}
\newtheorem{remark}{Remark}
\newtheorem{definition}{Definition}
\newtheorem{assumption}{Assumption}

\begin{document}
%
\title{Blind Beamforming for Coverage Enhancement with Intelligent Reflecting Surface}
\author{
	\IEEEauthorblockN{
	Fan Xu, \IEEEmembership{Member,~IEEE}, Jiawei Yao, \IEEEmembership{Student Member,~IEEE}, Wenhai Lai, \IEEEmembership{Student Member,~IEEE},\\
    Kaiming Shen, \IEEEmembership{Senior Member,~IEEE},
	Xin Li, Xin Chen, and Zhi-Quan Luo, \IEEEmembership{Fellow,~IEEE}
} 
\thanks{
Manuscript was accepted to IEEE Transactions on Wireless Communications on 11 July, 2024. 
 (\emph{Corresponding author: Kaiming Shen}.)

Fan Xu is with Peng Cheng Laboratory, Shenzhen (e-mail: xuf02@pcl.ac.cn).

Jiawei Yao, Wenhai Lai, and Kaiming Shen are with the School of Science and Engineering, The Chinese University of Hong Kong, Shenzhen (e-mail: jiaweiyao@link.cuhk.edu.cn; wenhailai@link.cuhk.edu.cn; shenkaiming@cuhk.edu.cn).

Xin Li and Xin Chen are with Huawei Technologies (e-mail: razor.lixin@huawei.com; chenxin@huawei.com).

Zhi-Quan Luo is with The Chinese University of Hong Kong, Shenzhen, with Shenzhen Research Institute of Big Data, China, and with Peng Cheng Laboratory, Shenzhen (e-mail: luozq@cuhk.edu.cn).
}
}


%


\maketitle

\begin{abstract}
Conventional policy for configuring an intelligent reflecting surface (IRS) typically requires channel state information (CSI), thus incurring substantial overhead costs and facing incompatibility with the current network protocols. This paper proposes a blind beamforming strategy in the absence of CSI, aiming to boost the minimum signal-to-noise ratio (SNR) among all the receiver positions, namely the coverage enhancement. Although some existing works already consider the IRS-assisted coverage enhancement without CSI, they assume certain position-channel models through which the channels can be recovered from the geographic locations. In contrast, our approach solely relies on the received signal power data, not assuming any position-channel model. We examine the achievability and converse of the proposed blind beamforming method. If the IRS has $N$ reflective elements and there are $U$ receiver positions, then our method guarantees the minimum SNR of $\Omega(N^2/U)$---which is fairly close to the upper bound $O(N+N^2\sqrt{\ln (NU)}/\sqrt[4]{U})$. Aside from the simulation results, we justify the practical use of blind beamforming in a field test at 2.6 GHz. According to the real-world experiment, the proposed blind beamforming method boosts the minimum SNR across seven random positions in a conference room by 18.22 dB, while the position-based method yields a boost of 12.08 dB.
\end{abstract}


\section{Introduction}\label{sec:overview}

\begin{figure}[!t]
\centering
\includegraphics[scale=0.7]{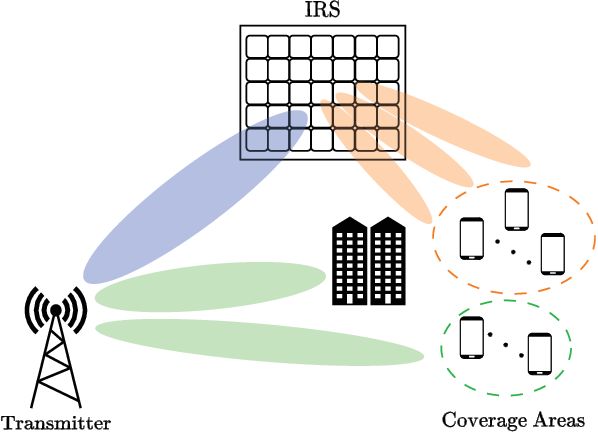}
\caption{IRS-assisted coverage enhancement.}\label{Fig scene graph}
\end{figure}

Intelligent reflecting surface (IRS) is an emerging wireless device that manipulates the reflected signals via phase shifting; it has attracted considerable research interest over the past few years, for it provides a low-cost and energy-efficient way of improving the wireless environment. This work focuses on the use of IRS in the coverage enhancement as shown in Fig.~\ref{Fig scene graph}. Specifically, we seek the optimal phase shifts of the different reflective elements (REs) of the IRS in order to maximize the minimum signal-to-noise (SNR) among all the possible receiver positions.
Differing from the previous attempts that either require the channel state information (CSI) \cite{Kassem2023,AlHilo2022,Pan2021,Yan2023,Sun2023} or assume certain position-channel models \cite{Lu2021,Ma2022,Mahbub2022}, the proposed \emph{blind beamforming} strategy does not entail any input or model other than a dataset of the received signal power, and hence it can be readily implemented in a plug-and-play fashion, as demonstrated by our field test at 2.6 GHz.

The idea of blind beamforming itself appears quite counter-intuitive. After all, how is it possible to address the optimization problem of the IRS beamforming even without knowing the channel coefficients in the problem? It turns out that the notion of blind beamforming is analogous to Nesterov's zero-order black box optimization \cite{Nesterov_book} and has already been explored in two recent works
\cite{Arun2020,blind_beamforming_twc} for a single receiver at the fixed position. Their main results stem from the fundamental fact that maximizing the conditional sample mean (CSM) of the received signal power for every RE is equivalent to aligning the corresponding reflected channel with the direct channel. Hence, it suffices to look at the data of the received signal power in order to maximize the SNR. However, it is quite difficult to extend the above result to multiple receiver positions, since the phase shifts optimized by the CSM method for a particular position can be a very bad solution for another position. To resolve such disagreement, we suggest a natural idea of majority voting: let every position vote for its favored phase shifts (as obtained from the CSM) across all the REs, and then for each RE select the phase shift that receives the most votes, which is referred to as 
\emph{majority-voting conditional sample mean (MV-CSM)} method. Moreover, \cite{Sun2023} and \cite{Yan2023a} suggest using the received signal power to acquire the CSI and then optimizing phase shifts for the IRS; this channel estimation approach is an important benchmark that our blind beamforming method will be compared with.

Our main result lies in the characterization of the performance limit of the MV-CSM method. Suppose that there are a total of $U$ randomly distributed receiver positions. For an IRS with $N$ REs, we show that the MV-CSM method achieves the following minimum SNR among the $U$ positions:
$$
\mathrm{SNR}_{\min}=\Omega\bigg(\frac{N^2}{U}\bigg)\;\;\text{with high probability (w.h.p.)},
$$
i.e., there exist a positive factor $c>0$ and a pair of positive integers $(N_0,U_0)$ such that $\mathrm{SNR}_{\min} \ge c N^2/U$ for any $N\ge N_0$ and any $U\ge U_0$. Regarding the converse, we show that the optimal performance under some mild conditions is bounded from above as
$$
\mathrm{SNR}_{\min}=O\bigg(N+\frac{N^2\sqrt{\ln (NU)}}{\sqrt[4]{U}}\bigg)\;\;\text{w.h.p.},
$$
i.e., there exist a positive factor $c>0$ and a pair of positive integers $(N_0,U_0)$ such that $\mathrm{SNR}_{\min} \le c N+c{N^2\sqrt{\ln (NU)}}/\sqrt[4]{U}$ for any $N\ge N_0$ and any $U\ge U_0$.

The use of IRS for the coverage enhancement purpose has been extensively studied in the literature to date. Considering the vehicular communications, \cite{Dhruvakumar2022} shows that the IRS deployment at a near-road position can benefit the signal reception along the entire road afar, while \cite{Javed2022} further shows that placing the IRS at the crossroad can boost the signal coverage for the present city block. The above two works however only consider the effectiveness of the IRS-assisted coverage enhancement, without detailed discussion on how the phase shifts of the IRS are optimally coordinated in order to maximize the performance gain. The authors of \cite{Kassem2023} propose a heuristic algorithm for optimizing the phase shifts of the IRS and also suggest dividing the IRS into multiple sub-IRSs each dedicated to one particular receiver position, under the assumption that the perfect CSI is known \emph{a priori}. The IRS beamforming problem is considered in \cite{AlHilo2022} in conjunction with the spectrum allocation, aiming to maximize the minimum rate at the various receiver positions. Its main idea is to optimize the phase shift for one RE at a time; this approach requires the CSI as well. For the same problem setup, \cite{Yan2023} advocates a gradient ascent approach. Also assuming the access to the CSI, \cite{Pan2021} pursues a sum-of-weighted-rates maximization across multiple receiver positions by means of the successive convex approximation. Our objective is most closely related to that in \cite{AlHilo2022,Yan2023,Sun2023}, i.e., the phase shifts of IRS are optimized to maximize the minimum SNR among a number of given receiver positions. However, unlike any of the above works, the present study does not assume the availability of CSI, and does not require channel acquisition either.

 
Actually, a line of previous attempts \cite{Lu2021,Ma2022,Mahbub2022} already consider the IRS beamforming for the coverage enhancement without CSI. Nevertheless, \cite{Lu2021,Ma2022,Mahbub2022} instead require the geographic location information, and particularly assume certain position-channel models through which the channels can be implicitly determined based on where the transmitter, receiver, and IRS are located. By virtue of this position-based channel model, the max-min SNR problem can be considered for a continuous position within the area, whereas the aforementioned model-free references \cite{Kassem2023,AlHilo2022,Yan2023,Pan2021} as well as our work is restricted to a finite number of discrete positions. However, the position-based approach in \cite{Lu2021,Ma2022,Mahbub2022} is valid only for the far-field and line-of-sight (LoS) scenarios. According to our field test, the state-of-the-art position-based method in \cite{Ma2022} is much inferior to the proposed blind beamforming method even for the transmissions inside a large empty room. Beam training algorithms \cite{Wang2022,Liu2023,An2022} constitute another existing approach to the IRS beamforming in the absence of channel knowledge. Their idea is fairly straightforward: try out a sequence of codewords (each corresponding to a possible array of phase shifts for the IRS) and then pick the best. This paper shows analytically that the proposed algorithm can strictly outperform the beam training method; this advantage is further demonstrated in field tests and simulations.
In the realm of information theory, \cite{Zhi2023} derives a closed-form achievable rate for the multi-user IRS-assisted system with imperfect CSI, \cite{Zhu2023} proposes an efficient channel acquisition method for the IRS, and \cite{Cheng2021a} examines the channel capacity of an IRS-assisted system in which both the transmitter and the IRS can send their independent messages toward the receiver; these works either assume the CSI available or aim to acquire the CSI.

The rest of this paper is organized as follows. Section \ref{sec:model} describes the IRS-assisted coverage enhancement problem. Section \ref{section CSM} reviews the blind beamforming method in \cite{Arun2020,blind_beamforming_twc} designed for a single receiver whose location is fixed. Section \ref{sec blind beamforming} proposes an extended blind beamforming method called MV-CSM to account for multiple receiver positions, followed by the achievability and converse analyses. Section \ref{section experiment} shows the numerical results of our field test and simulation. Finally, Section \ref{sec:conclucion} concludes this paper.

The notation here and throughout is summarized as follows. We use the bold lower-case letter to denote a vector, the bold upper-case letter a matrix, and the calligraphy letter a set. For a complex number $a$, denote by $\angle a\in(-\pi,\pi]$ its principal argument. The Bachmann-Landau notation is used extensively in this paper: write $f(n) = O(g(n))$ if there exists a positive factor $c$ and a positive integer $n_0$ such that $f(n)\le cg(n)$ as $n\ge n_0$, write $f(n)=\Omega(g(n))$ if $g(n)= O(f(n))$, write $f(n)=\Theta(f(n))$ if $f(n) = O(g(n))$ and $f(n)=\Omega(g(n))$ both hold, write $f(n)=o(g(n))$ if for any $c>0$ there exists a positive integer $n_0$ such that $f(n)\le cg(n)$ as $n\ge n_0$, and write $f(n)=\omega(g(n))$ if for any $c>0$ there exists a positive integer $n_0$ such that $f(n)\ge cg(n)$ as $n\ge n_0$. Suppose that the variable $X>0$ depends on another variable $Y>0$; write $X\sim Y$ if $\lim_{Y\rightarrow\infty}X/Y=1$, i.e., they are asymptotically equal. For an event $\mathcal{A}$, denote by $\mathbb P\{\mathcal A\}$ the probability of the event.
Moreover, we use $\mathcal U(a,b)$ to denote the uniform distribution on the closed interval $[a,b]$, $\mathcal B(n,p)$ the binomial distribution with $n$ trials and $p$ being the success probability of each trial, and $\mathcal{CN}(\boldsymbol{\mu},\mathbf{\Gamma})$ the complex Gaussian distribution with the mean $\boldsymbol{\mu}$ and the covariance matrix $\mathbf{\Gamma}$.

\section{Problem Statement}
\label{sec:model}

Consider an IRS-assisted downlink network. Assume that the IRS consists of $N$ REs, each controlled by the corresponding phase shift $\theta_n, n=1,\ldots,N$. Assume also that the user terminal is likely to appear at any of the given $U$ receiver positions. Denote by $h_{u,0}\in\mathbb C$ the direct channel from base-station to the receiver position $u=1,\ldots,U$; denote by $h_{u,n}\in\mathbb C$ the reflected channel from the base-station to the RE $n$ and ultimately to the receiver position $u$. For the transmit signal $X\in\mathbb C$ and the i.i.d. complex Gaussian noise $Z_u\sim\mathcal{CN}(0,\sigma^2)$ at position $u$, the received signal $Y_u\in\mathbb C$ at position $u$ is given by
\begin{equation}
\label{Y:U-IRS}
    Y_u = \left(h_{u,0}+\sum_{n=1}^N h_{u,n}e^{j\theta_{n}}\right)X+Z_u.
\end{equation}
With the transmit power $P=\mathbb E[|X|^2]$, the SNR at position $u$ can be computed as
\begin{equation}
    \mathrm{SNR}_u=
    \left|h_{u,0}+\sum_{n=1}^N h_{u,n}e^{j\theta_{n}}\right|^2\frac{P}{\sigma^2}.\label{eqn SNR u}
\end{equation}
Furthermore, in the practical implementation, the choice of each phase shift $\theta_n$ is restricted to a prescribed discrete set
\begin{equation}
    \Phi_K = \left\{0,\frac{2\pi}{K},2\times \frac{2\pi}{K},\ldots,(K-1)\times \frac{2\pi}{K}\right\}
\end{equation}
for a positive integer $K\ge2$.

Suppose that the users are densely distributed throughout the target area; then maximizing the minimum rate among these users amounts to the coverage enhancement problem. Thus, we seek the optimal phase shifts $(\theta_1,\ldots,\theta_N)$ to maximize the minimum SNR among the $U$ possible receiver positions:
\begin{subequations}\label{eqn problem}
\begin{align}
\underset{ (\theta_1,\ldots,\theta_N) }{\text{maximize}}\quad & \mathrm{SNR}_\text{min}\triangleq\min_{u}\Big\{\mathrm{SNR}_u\Big\}\label{eqn problem1}\\ 
\textrm{subject to}\quad &\theta_{n}\in\Phi_K,\;\text{for}\;n=1,\ldots,N.\label{eqn problem3}
\end{align}
\end{subequations}
Note that the channels $\{h_{u,0},h_{u,n}\}$ are not available in the above problem. 

One may wonder why not first estimate the channels and then solve the problem in \eqref{eqn problem}. There are several issues that prevent us from doing that. First, each reflected channel $h_{u,n}$ alone is very weak, so it is difficult to recover the reflected channels precisely. Second, the channel acquisition can be costly when the IRS comprises a large number of REs. Third, 
the existing network protocols must be modified in order to support the additional channel estimation for the IRS. Due to the above concerns, none of the existing IRS prototypes \cite{Arun2020,Dai2020,Liyanage2023} adopts channel estimation.

\section{Blind Beamforming for One Position}\label{section CSM}

We start by reviewing how the phase shifts are optimized blindly for a single receiver position (i.e., when $U=1$) as considered in \cite{Arun2020,blind_beamforming_twc}. For ease of notation, we drop the user index $u=1$ throughout this section.

Notice that the problem \eqref{eqn problem} with $U=1$ reduces to a trivial one when (i) CSI is available and (ii) $K\rightarrow\infty$. In this case, it is optimal to align each reflected channel $h_n$ with the direct channel $h_0$ by letting $\theta_n=(\angle h_0 - \angle h_n) \bmod 2\pi$. When $K$ is finite, a natural idea is to round the above continuous solution to the closest point in the discrete set $\Phi_K$, namely the \emph{closest point projection (CPP)}:
\begin{align}
\label{CPP}
&\theta^\text{CPP}_n=\arg\min_{\phi\in\Phi_K}\left|\angle\left(\frac{h_ne^{j\phi}}{h_0}\right)\right|.
\end{align}
A surprising result from \cite{Arun2020,blind_beamforming_twc} is that the CPP solution can be recovered without CSI, as specified in the following.

First, generate $T$ random samples of the phase shift array uniformly and independently, sample $t$ denoted by $(\theta_1^{(t)},\ldots,\theta_N^{(t)})$, for $t=1,\ldots,T$. With respect to each sample $t$, measure the received signal power $|Y^{(t)}|^2$. Furthermore, for each $n=1,\ldots,N$ and each $k=0,\ldots,K-1$, a subset of samples, $\mathcal G_{n,k}\subset\{1,\ldots,T\}$, is defined to be
\begin{align}
\mathcal{G}_{n,k}=\left\{ t:\theta^{(t)}_{n}=\frac{2\pi k}{K}\right\}.\label{eqn set of sample indices}
\end{align}
The CSM of the received signal power is now computed with respect to each $\mathcal{G}_{n,k}$, i.e.,
\begin{equation}
\label{cond_expectation}
\widehat{\mathbb E}\left[|Y|^2\Big|\theta_{n}=\frac{2\pi k}{K}\right]=\frac{1}{|\mathcal{G}_{n,k}|} \sum\limits_{t \in \mathcal{G}_{n,k}} |Y^{(t)}|^2.
\end{equation}
Intuitively speaking, $\widehat{\mathbb E}\big[|Y|^2|\theta_{n}=\frac{2\pi k}{K}\big]$ reflects the average gain of letting $\theta_n=2\pi k/K$ when the rest phase shifts are random. CSM \cite{Arun2020,blind_beamforming_twc} decides each phase shift as 
\begin{align}
\label{CSM}
    \theta^{\textrm{CSM}}_n = \arg \max_{\phi\in\Phi_K} \widehat{\mathbb{E}}[|Y|^2|\theta_n=\phi].
\end{align} 
It turns out that CSM can recover the solution $\theta_n^\text{CPP}$ for $T$ sufficiently large, as stated in the following proposition.
\begin{proposition}[Theorem 2 and Remark 1 in \cite{blind_beamforming_twc}]
\label{prop:CSM}
When the sample size $T=\Omega(N^2(\log N)^3)$ and the area of each RE is fixed, the CSM method and the CPP method yield the same solution, i.e., $\theta^\text{CSM}_n=\theta^\text{CPP}_n$ for $n=1,\ldots,N$. As a result, the CSM method guarantees that
\begin{equation}
\cos^2(\frac{\pi}{K})\cdot f^\star\le \mathbb E[\mathrm{SNR}] \le f^\star,
\end{equation}
where $f^\star$ is the global optimum of \eqref{eqn problem}. Thus, if the number of phase shift choices $K>2$, 
then the SNR achieved by CSM grows quadratically with the number of REs $N$, i.e.,
\begin{equation}
\label{boost:single_IRS}
    \mathbb E\left[\mathrm{SNR}\right] = \rho\cdot\Theta(N^2)
\end{equation}
with the average reflection power gain
\begin{equation}
    \rho = \frac{\sum_{n=1}^N |h_n|^2}{N},
\end{equation}
where the expectation is taken over random samples.
\end{proposition}

We remark that the condition $K>2$ is necessary for CSM to guarantee a quadratic SNR boost in $N$. In other words, when $K=2$, the performance gain by CSM can be arbitrarily close to zero, as illustrated with an example in \cite[Remark 2]{blind_beamforming_twc}. We further intuitively explain why the assertion of Proposition \ref{prop:CSM} holds true. Consider the extreme case in which $N$ is finite while $T\rightarrow\infty$. The CSM value $\widehat{\mathbb E}\left[|Y|^2\big|\theta_{n}=\phi\right]$ would converge to the conditional expectation:
\begin{multline}
\mathbb E\left[|Y|^2\big|\theta_{n}=\phi\right] = 2P|h_{u,0}h_{u,n}|\cos(\angle h_{u,0}-\angle h_{u,n}-\phi)\\
+\sum^N_{n'=0}|h_{u,n'}|^2P+\sigma^2,
\end{multline}
where the expectation is taken over independent and uniformly distributed $\theta_{n'}\in\Phi_K$ ($n'\ne n$) and the complex Gaussian random variable $X$. 
Since the CSM method aims to maximize the left-hand side of the above equation, it would choose $\phi$ to minimize the gap between $\angle h_{u,0}$ and $h_{u,n}+\phi$ on the right-hand side; this operation can be recognized as the CPP method in \eqref{CPP}, and the result of Proposition \ref{prop:CSM} immediately follows. The analysis for the finite-$T$ case is far more complicated. The reader is referred to \cite{blind_beamforming_twc} for the complete proof. Next, we illustrate the procedure of the CSM method through a toy example.

\allowdisplaybreaks[4]

\begin{table}[t]
\centering
\caption{A toy example of applying CSM to an IRS-assisted network with $N=4$, $K=2$, and $T=6$.}
\label{table example CSM}
\small
\begin{tabular}{|c|c|c|}
\hline
$t$ & $(\theta_1,\theta_2,\theta_3,\theta_4)$     & $|Y|^2$ \\ \hline
\hline
1     & $(0,\pi,0,0)$     & 2.8          \\ \hline
2     & $(0,0,0,0)$       & 1.0          \\ \hline
3     & $(\pi,\pi,\pi,0)$ & 1.5      \\ \hline
4     & $(\pi,0,\pi,\pi)$ & 3.3      \\ \hline
5     & $(\pi,\pi,0,\pi)$ & 0.3      \\ \hline
6     & $(0,0,\pi,\pi)$   & 0.4         \\ \hline
\end{tabular}
\end{table}

\begin{example}
Consider the random samples in Table \ref{table example CSM}. Based on \eqref{cond_expectation}, the CSM with $\theta_1=0$ and the CSM with $\theta_1=\pi$ are computed as
\begin{align}
&\widehat{\mathbb E}\left[|Y|^2\Big|\theta_{1}=0\right]=\frac{2.8+1.0+0.4}{3}=1.4\notag
\end{align}
and
\begin{align}
&\widehat{\mathbb E}\left[|Y|^2\Big|\theta_{1}=\pi\right]=\frac{1.5+3.3+0.3}{3}=1.7.\notag
\end{align}
Thus, the CSM method lets $\theta_1=\pi$ according to \eqref{CSM}. The rest phase shifts can be decided similarly. The CSM solution is then $(\theta_1,\theta_2,\theta_3,\theta_4)=(\pi,0,\pi,0)$. It is worth mentioning that the above solution does not appear in the six random samples.
\end{example}

We conclude this section by discussing the practical aspect of the CSM method in the following two remarks.

\begin{remark}[CSM vs. Beam Training]\label{remark compare RMS}
Beam training method \cite{Wang2022,Liu2023,An2022} is another existing approach to the IRS beamforming without any channel knowledge. A common implementation of beam training is based on the uniformly generated codebook, which can be recognized as the so-called Random Max-Sampling (RMS) method in \cite{blind_beamforming_twc}, i.e., try out $T$ i.i.d. random samples of the IRS beamformers and then pick the best one. However, it is shown in \cite{blind_beamforming_twc} that the SNR boost by RMS is at most $\Theta(N\log T)$. Comparing this result with Proposition \ref{prop:CSM}, we conclude that CSM is asymptotically superior to RMS (or the uniform beam training) since the number of random samples, $T$, is typically polynomial in $N$.
\end{remark}

\begin{remark}[Training Cost of CSM]
From the network protocol point of view, the number of time slots should be equal to $T$, since each random sample corresponds to a transmitted symbol $X$ that occupies one time slot. Nevertheless, the time span of random samples is also affected by the hardware constraints in practice. In our prototype case, the phase shifts of the IRS device can only be updated 1 time every 1 second due to the PN circuit limit. In other words, after one random sample is taken, we need to wait 1 second till the next random sample. Thus, if we perform $T$ random samples in a row, then it would require $T$ seconds---which can be longer than $T$ time slots. In our field tests as shown in Section \ref{field tests}, we let $T=500$ so it takes around 8 minutes to complete random sampling. Since the IRS configuration in our problem scenario is considered in the long run (typically in hours), the above time cost is tolerable. Besides, there are also two possible ways to reduce the time cost of random sampling. First, the time period between two random samples could be exploited for data transmission. Second, the hardware implementation of the IRS can be further improved to allow the phase shifts to be switched much more frequently; there is actually huge room for improvement for our prototype machine.
\end{remark}

\section{Blind Beamforming for Multiple Positions}\label{sec blind beamforming}

\subsection{Proposed MV-CSM Method}\label{sec blind beamforming proposed MVCSM}

Our goal here is to extend the above CSM method to the general $U\ge2$-position case in \eqref{eqn problem3} with provable performance.
The proposed extension is fairly simple: as its name MV-CSM implies, the proposed method just lets the positions vote for their favored phase shifts and then decides each $\theta_n$ according to the majority rule.
Specifically, with the CSM solution of each position $u$ denoted by $(\theta^\text{CSM}_{u,1},\ldots,\theta^\text{CSM}_{u,N})$, the MV-CSM method chooses each phase shift as
\begin{equation}
\label{MV-CSM}
    \theta^\text{MV-CSM}_n = \arg\max_{\phi\in\Phi_K}\sum^U_{u=1}\mathbbm 1\{\theta^\text{CSM}_{u,n}=\phi\}.
\end{equation}
If there exists a tie then we break it randomly. (In other words, if more than one phase shift candidate receive the most votes, then choose any one of them randomly.)

We point out that the above voting procedure can be alternatively interpreted as maximizing the CSM of a type of utility rather than the received signal power. When the phase shift $\theta_n$ is set to some $\phi\in\Phi_K$, position $u$ is said to be \emph{satisfied} if $\phi$ is exactly its favored phase shift for RE $n$. Accordingly, the utility of $\theta_n$ is defined to be the number of satisfied positions under this phase shift choice. It can be seen that the CSM of this satisfactory utility is more aggressive than the CSM of the received signal power; $\theta_n$ receives one credit from position $u$ only if it matches the best phase shift choice for $u$.

Moreover, observe that the main workload of the MV-CSM method lies in performing CSM for each individual position. Since CSM runs in $O(NT)$ time \cite{Arun2020,blind_beamforming_twc}, the overall time complexity of MV-CSM is $O(NTU)$, which amounts to $O(N^2(\log N)^3U)$ if the number of samples is set to the lower bound in Proposition \ref{prop:CSM}. 

\subsection{Achievability Analysis}
\label{section performance achievable}

Our analyses here and throughout are based on the following mild conditions:

\begin{assumption}
\label{assumption:1}
    The channel phases $\angle h_{u,n}$ are uniformly distributed on $[0,2\pi)$ across $u=1,\ldots,U$ and also across $n=0,1,\ldots,N$. The reflected channels toward the same receiver position $u$ are of the same magnitude $c_u>0$, i.e., i.e., $|h_{u,n}|=c_u$ for all $n=1,\ldots,N$.
\end{assumption}

The following theorem and corollary give the asymptotic performance guaranteed by the MV-CSM method.
\begin{theorem}
\label{prop achievable}
Assume that $N=\omega(U^2)$, $T=\Omega(N^2(\ln NU)^3+N^2U(\ln NU))$, and the area of each RE is fixed. Assume also that the binary random sampling is used, i.e., each $\theta_n$ is uniformly and independently selected from the set $\{0,\pi\}$ for each random sample. Then under Assumption \ref{assumption:1} the MV-CSM method yields
\begin{align}\label{eqn achievable SNR}
\textrm{SNR}_u=\frac{4Pc_u^2}{\sigma^2\pi^2}\cdot\Omega\bigg(\frac{N^2}{U}\bigg) \quad\text{w.h.p.}
\end{align}
at each receiver position $u$.
\end{theorem}
\begin{IEEEproof}
    See Appendix A.
\end{IEEEproof}

\begin{corollary}\label{corollary achievable min}
Under the same conditions as stated in Theorem \ref{prop achievable},  the MV-CSM method yields
\begin{equation}
\label{eqn achievable minSNR}
\textrm{SNR}_\textrm{min}=\frac{4Pc_\textrm{min}^2}{\sigma^2\pi^2}\cdot\Omega\bigg(\frac{N^2}{U}\bigg)\quad\text{w.h.p.},
\end{equation}
where $c_\textrm{min}=\min_u c_u$.
\end{corollary}

Of particular interest is the contrast between Proposition \ref{prop:CSM} and Theorem \ref{prop achievable}. The former requires $K>2$ to prove the performance of CSM, while the latter requires $K=2$ to prove the performance of MV-CSM. It is much more difficult to analyze the performance of MV-CSM. Of course, since $K$ is often a power of 2 in practice, we can always reduce the $K$-ary beamforming to the binary beamforming by limiting each $\theta_n$ to $\{0,\pi\}$, and thereby have the performance guarantee in the above theorem and corollary.

We now explain the results of Theorem \ref{prop achievable} and Corollary \ref{corollary achievable min} intuitively. Recall that a particular position $u$ is satisfied with RE $n$ if $\theta_n$ happens to be its CSM solution for RE $n$. Asymptotically, the proof of Theorem \ref{prop achievable} shows that each position $u$ is satisfied with at least $N/2+\Omega(N/\sqrt{U})$ REs; in other words, it is unsatisfied with at most $N/2-\Omega(N/\sqrt{U})$ REs. After these reflected channels cancel out with each other, there remain at least $\Omega(N/\sqrt{U})$ REs satisfactory to position $u$. The results of Theorem \ref{prop achievable} and Corollary \ref{corollary achievable min} then immediately follow. In particular, the SNR performance is inversely proportional to $U$, because it is increasingly difficult to reach a consensus on $\theta_n$ among the different positions when more positions join in the vote.

Importantly, the proposed method is based on the long-term statistics that capture the large-scale features of the wireless environment, so a particular position in the target area should still be well improved even if it is not directly optimized by the MV-CSM, as long as its nearby positions have been optimized by the MV-CSM. Furthermore, the following three remarks delve deeply into Theorem \ref{prop achievable}.

\begin{remark}[Why Binary Random Sampling?]
    In the above theorem, we restrict the choice of each $\theta_n$ to the binary set $\{0,\pi\}$. In other words, even when the IRS can provide more phase shift choices $\{0,\frac{2\pi}{K},2\times \frac{2\pi}{K},\ldots,(K-1)\times \frac{2\pi}{K}\}$, we choose to use the subset $\{0,\pi\}$. But what is the benefit of limiting $\theta_n$ to $\{0,\pi\}$? This setting is critical to Lemma \ref{lemma achievable xi_u} shown in Appendix A and thereby can greatly facilitate the performance analysis. But one could have used a larger range of phase shift choices when the IRS has $K>2$ reflective elements and then the proposed algorithm continues to work. The only issue with using $K>2$ is that it is difficult to verify the performance of the proposed algorithm since the important tool Lemma \ref{lemma achievable xi_u} then becomes invalid.
\end{remark}

\begin{remark}[Why Fixed Area for Each RE?]
Basically, the received signal power with the assistance of IRS is given by \cite{Tang2022}:
$$
\text{received signal power} \propto N^2SG,
$$
where $S$ is the area of each RE and $G$ is the scattering gain. Note that the product $SG$ can be interpreted as the overall signal power reflected by each RE. Thus, if the value of $SG$ is fixed, then the received signal power (or, equivalently, the SNR) grows quadratically with $N$, as considered in many previous works \cite{blind_beamforming_twc,Han2020,Han2022,Wu2021b}. The conclusion in our paper is also based on the assumption that $SG$ is fixed. Nevertheless, such quadratic growth cannot persist forever because of the energy conservation. This hunch stems from physics, but can also be explained based on the above model of the received signal power. Assume that the total area of the IRS equals $A$. Now we fix $A$ and let $N$ tend to infinity. Evidently, the area of each RE would shrink as
$$
S \propto \frac{1}{N}.
$$
Moreover, as shown in \cite{Richards2014,Tang2022}, the scattering gain $G$ is linear in $S$ according to the electromagnetic theory, i.e.,
$$
G \propto S.
$$
As a result, when $A$ is fixed, increasing $N$ cannot bring any boost because $SG$ would decrease quadratically at the same time. In other words, the quadratic boost can be achieved only when $S$ is fixed, so that $A$ would grow linearly with $N$---but this does not hold forever because of the space limit. We clarify that our analysis is under the condition that the area per RE is fixed and thus $SG$ is fixed. This condition can be secured so long as the IRS area $A$ is far smaller than the free space---which is often the case in practice.
\end{remark}

\begin{remark}[Extension for Active IRS]
    Furthermore, following the previous work \cite{Zhi2022a}, we consider the active IRS case in which the IRS can amplify the incident signals with a factor of $\rho\in(0,\rho_{\max}]$, where $\rho_{\max}$ is the power budget at the IRS. Recall that the proposed algorithm can enhance the SNR of every user (as discussed under Theorem \ref{prop achievable}), so the reflected channels are constructive in boosting the overall channel strength for every user, and thus scaling up all the reflected channels simultaneously can enhance the SNR further. As such, it is optimal to set $\rho=\rho_{\max}$ when the proposed algorithm is used for an active IRS. Nevertheless, as discussed in \cite{Zhi2022a}, the reflection at the IRS can incur extra noise, namely the IRS noise. If the IRS noise cannot be neglected, then it is no longer easy to decide the optimal $\rho$ for our algorithm. This difficult problem can be a future research direction.
\end{remark}

\begin{table*}[t]
\renewcommand{\arraystretch}{1.4}
\setlength{\tabcolsep}{3pt}
\centering
\caption{Computational Complexities of the Different Algorithms for the IRS Beamforming}\label{table complexity}
\small
\begin{tabular}{|c|c|ccccc|}
\hline
\multirow{3}{*}{MV-CSM} & \multirow{3}{*}{\begin{tabular}[c]{@{}c@{}}Position-Based\\ Method \cite{Ma2022} \end{tabular}}                                & \multicolumn{5}{c|}{Channel Estimation-based Methods}                                                                                                                            \\ \cline{3-7} 
&                                                     & \multicolumn{3}{c|}{Channel Estimation}                                                             & \multicolumn{2}{c|}{Phase Shift Optimization}                              \\ \cline{3-7} 
&                                                     & \multicolumn{1}{c|}{\begin{tabular}[c]{@{}c@{}}DFT \cite{channel_est_DFT} \end{tabular}}       & \multicolumn{1}{c|}{\begin{tabular}[c]{@{}c@{}}Autocorrelation \cite{Yan2023a} \end{tabular}}           & \multicolumn{1}{c|}{\begin{tabular}[c]{@{}c@{}}Neural Network \cite{Sun2023}\end{tabular}}      & \multicolumn{1}{c|}{\begin{tabular}[c]{@{}c@{}}SDR  \cite{Sun2023} \end{tabular}}          & Gradient Ascent \cite{Yan2023}                                    \\ \hline
$O(NUT)$                & $O((N_x^{6.5}\!+\!N_z^{6.5})\log(\frac{1}{\epsilon})I)$ & \multicolumn{1}{c|}{$O(UN^3)$} & \multicolumn{1}{c|}{$O(UN^{4.5}I)$} & \multicolumn{1}{c|}{$O(UND)$} & \multicolumn{1}{c|}{$O(N^{3.5})$} & $O((NU\!+\!\log(\frac{1}{\epsilon})\!+\!NK)I)$ \\ \hline
\end{tabular}
    \begin{tablenotes}
      \small
      \item Note: $I$ is the number of iterations for an iterative algorithm to converge, $\epsilon>0$ is the accuracy of the interior-point optimization,  $N_x$ is the number of columns of the RE array, $N_z$ is the number of rows of the RE array, and $D$ is the training dateset size for deep learning.
    \end{tablenotes}
\end{table*}

\subsection{Converse Analysis}
\label{section:converse}

Our discussion of the converse is restricted to a class of ``good'' algorithms as defined below:

\begin{definition}
\label{defiition:good_alg}
An IRS beamforming algorithm is said to be good if it enables the IRS to improve the SNR for every position $u=1,\ldots,U$.
\end{definition}

Notice that an IRS beamforming algorithm that enhances the minimum SNR is not necessarily good. But the MV-CSM method must be a good algorithm according to Theorem \ref{prop achievable}. The following theorem bounds the minimum SNR from above for any good algorithm.

\begin{theorem}\label{prop converse}
Under Assumption \ref{assumption:1}, the minimum SNR achieved by any good IRS beamforming algorithm satisfies
\begin{align}
&\textrm{SNR}_\textrm{min}= \frac{P\eta^2 c_{\textrm{max}}^2}{\sigma^2}\cdot O\left(N+\frac{N^2\sqrt{\ln (NU)}}{\sqrt[4]{U}}\right)\quad\text{w.h.p.},
\label{eqn upper bound}
\end{align}
where $c_{\max}=\max_{u}|c_u|$ and $\eta= \max_u|h_{u,0}|/\min_u|h_{u,0}|$.
\end{theorem}
\begin{IEEEproof}
    See Appendix B.
\end{IEEEproof}

The above converse applies to the MV-CSM method because it is a good IRS beamforming algorithm. Thus, combining the inner bound in Theorem \ref{prop achievable} and the outer bound in Theorem \ref{prop converse} gives an approximation ratio for the MV-CSM method, as stated in the following corollary.

\begin{corollary}
Consider the same condition as stated in Theorem \ref{prop achievable}. The minimum SNR achieved by MV-CSM, denoted as $\mathrm{SNR}_\text{min}$, and the global optimum $\mathrm{SNR}^\star_\text{min}$ for a good IRS beamforming algorithm satisfy
\begin{equation}
\frac{\mathrm{SNR}^\star_\text{min}}{\mathrm{SNR}_\text{min}} = O\Big({U^{\frac34}}\sqrt{\ln (NU)}\Big)\quad\text{w.h.p.}
\end{equation}
\end{corollary}

\subsection{Partitioning Conditional Sample Mean (P-CSM)}
\label{subsec:P-CSM}

For the comparison purpose, we also examine the performance limit of a baseline method called the partitioning conditional sample mean (P-CSM). Its central idea is to partition the IRS into multiple blocks and then optimize each block for one particular receiver position; this IRS partitioning idea is widely considered in the literature \cite{You2022,Kassem2023,Aldababsa2021}.

In our case, we assume that $N$ is much greater than $U$ so that $N$ can be approximated as a multiple of $U$. The IRS is equally partitioned into $U$ blocks; the phase shifts of each block $u$ are optimized by the CSM method according to the received signal power at position $u$. The performance limit of P-CSM is given in the following theorem.

\begin{theorem}\label{prop achievable P-CSM}
Consider the same conditions as in Theorem \ref{prop achievable} except that we now let $N=\omega(U^3)$. Then under Assumption \ref{assumption:1}, the P-CSM method yields
\begin{align}\label{eqn achievable SNR P-CSM}
\mathrm{SNR}_u = \frac{4Pc^2_u}{\sigma^2\pi^2}\cdot\Omega\bigg(\frac{N^2}{U^2}\bigg)\quad\text{w.h.p.}
\end{align}
for every position $u=1,\ldots,U$. Moreover, the minimum SNR achieved by the P-CSM method satisfies
\begin{align}
\label{eqn achievable minSNR:converse}
\mathrm{SNR}_{\min} = \frac{4Pc^2_{\min}}{\sigma^2\pi^2}\cdot\Omega\bigg(\frac{N^2}{U^2}\bigg)\quad\text{w.h.p.},
\end{align}
where $c_{\min}=\min_u c_u$.
\end{theorem}
\begin{IEEEproof}
    See Appendix C.
\end{IEEEproof}

As the contrast between \eqref{eqn achievable minSNR} and \eqref{eqn achievable minSNR:converse} indicates, MV-CSM achieves a higher SNR boost than MV-CSM does in general.

We now further compare the computational complexities of the proposed algorithm and other existing algorithms.
The complexity of MV-CSM is $O(NUT)$ as already analyzed in Section \ref{sec blind beamforming proposed MVCSM}, and it is easy to verify that the complexity of P-CSM is $O(NT)$ according to \cite{blind_beamforming_twc}.  Table \ref{table complexity} summarizes the computational complexities of the different algorithms (either for channel estimation or for phase shift optimization).

\subsection{Random Sampling vs. Channel Estimation}

As stated in Theorem \ref{prop achievable}, the proposed blind beamforming algorithm, MV-CSM, has provable good performance so long as $T=\Omega(N^2(\ln NU)^3+N^2U(\ln NU))$. In contrast, the existing works \cite{Mishra2019,Jensen2020,Wang2020} in the literature claim that sending $N$ pilot symbols suffices to estimate all the channels of an IRS system. One may thus conclude that the channel estimation approach is actually more efficient. But we argue that this is a misconception.

First of all, we would like to clarify that $T=\Theta(N)$ is just a \emph{lower bound} on the number of pilot symbols for the channel estimation method to work. Specifically, there are $N$ unknown channel coefficients and thus it requires at least $N$ equations (each corresponding to a pilot symbol) to solve the linear system, but this lower bound of $T=\Theta(N)$ cannot guarantee the performance of channel estimation based method at all in the presence of additive noise. In contrast, the proposed bound $T=\Theta(N^2(\log N)^3)$ is an \emph{upper bound} on the number of random samples for CSM. In other words, it requires at most $T=\Theta(N^2(\log N)^3)$ random samples for the proposed algorithm to achieve the performance as promised in the proposition/theorem. Notice that a lower-bound complexity of Method A being smaller than an upper-bound complexity of Method B does NOT suggest that A is more efficient than B, especially when the performance of A cannot be guaranteed with such lower-bound complexity.

Besides, if $T$ falls below the upper bound $\Theta(N^2(\log N)^3)$, the proposed MV-CSM method continues to work only that its performance cannot be guaranteed by our current theory anymore. In contrast, if there are fewer than $N$ pilot symbols, then channel estimation based method cannot work at all (unless under certain sparsity conditions). We numerically demonstrate this point in Section \ref{section experiment}.

Last, from a practical implementation perspective, the channel estimation methods require the IRS devise to read the received symbol $Y\in\mathbb C$ from the communication chip of the receiver device---which is not supported by the current network protocol. In contrast, CSM only entails the received signal power information, which can be readily obtained from most receiver devices on the market. Moreover, many existing channel estimation methods \cite{Nadeem2020,Zheng2020b,Jensen2020,Wang2020} require setting each phase shift according to the DFT matrix, but this can violate the discrete phase shift constraint, e.g., they do not work for the case where each phase shift is limited to $\{0,\pi\}$.

\section{Experiments}\label{section experiment}

\subsection{Competitor Algorithms}

We summarize the proposed method and the benchmark methods. First, we consider two baseline cases that do not optimize phase shifts for the IRS or do not deploy the IRS:
\begin{itemize}
\item Zero Phase Shift: it fixes the phase shift of each RE at zero.
\item Without IRS: this is the baseline case without the IRS deployment.
\end{itemize}
We then consider blind beamforming in the absence of CSI:
\begin{itemize}
\item MV-CSM: this is our proposed method as shown in \eqref{MV-CSM}.
\item P-CSM: this is a naive extension of the CSM method \cite{blind_beamforming_twc} to the multi-user case as described in Section \ref{subsec:P-CSM}.
\item Position-Based Method \cite{Ma2022}: it optimizes phase shifts based on the geometric positions without CSI.
\item Random Max-Sampling (RMS) \cite{blind_beamforming_twc}: this is the beam training method using a uniform codebook.
\item Euclidean Max-Sampling (EMS) \cite{An2022}: it optimizes phase shifts based on the Euclidean distances.
\end{itemize}
Moreover, we consider the channel estimation based methods that first acquire CSI and then solve the IRS beamforming problem explicitly:
\begin{itemize}
    \item Neural + SDR \cite{Sun2023}: it uses neural network to acquire CSI and then optimizes phase shifts by SDR.
    \item Neural + Gradient \cite{Sun2023,Yan2023}: it uses neural network to acquire CSI and then optimizes phase shifts by gradient ascent.
    \item DFT + SDR \cite{channel_est_DFT,Sun2023}: it uses the DFT method to acquire CSI and then optimizes phase shifts by SDR.
    \item DFT + Gradient \cite{channel_est_DFT,Yan2023}: it uses the DFT method to acquire CSI and then optimizes phase shifts by gradient ascent.
    \item Autocorrelation + SDR \cite{Yan2023a,Sun2023}: it uses the autocorrelation of received signal to acquire CSI and then optimizes phase shifts by SDR.
\end{itemize}
We remark that the neural network method for channel estimation is adopted for field tests because it only requires the received signal power and thus can be easily implemented in our prototype system (although it is time-consuming). In contrast, the DFT method requires reading the received signal phase from the communication chip and also setting phase shifts according to the DFT matrix, so it is incompatible with our prototype system; as such, we consider the DFT method only in the simulations. Moreover, we only consider the autocorrelation method in simulation because it cannot finish in reasonable time in our field tests.

\begin{figure}
\centering
\subfigure[]
{
\includegraphics[width=6cm]{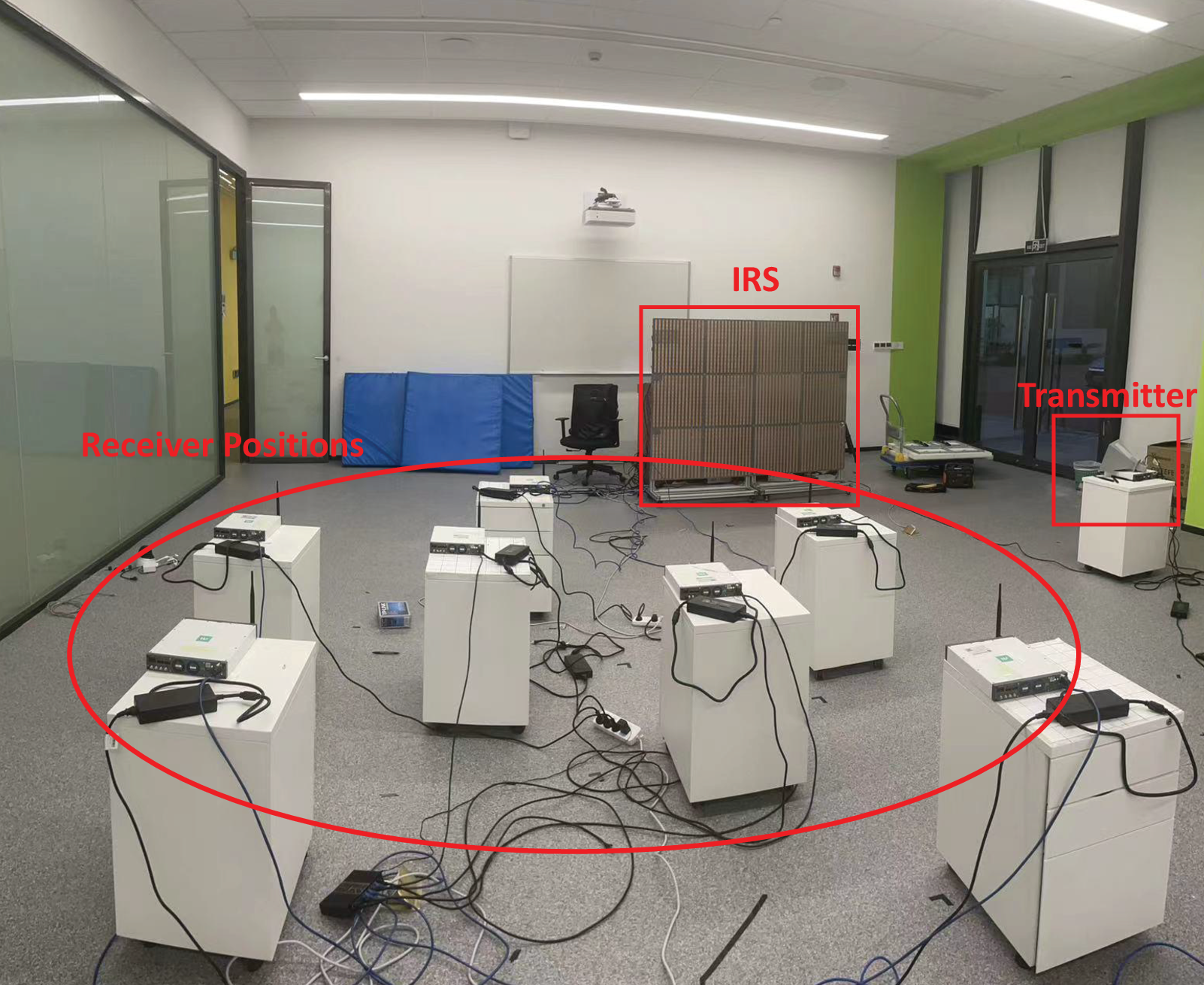}
}\\
\vspace{1em}
\subfigure[]
{
\includegraphics[width=6cm]{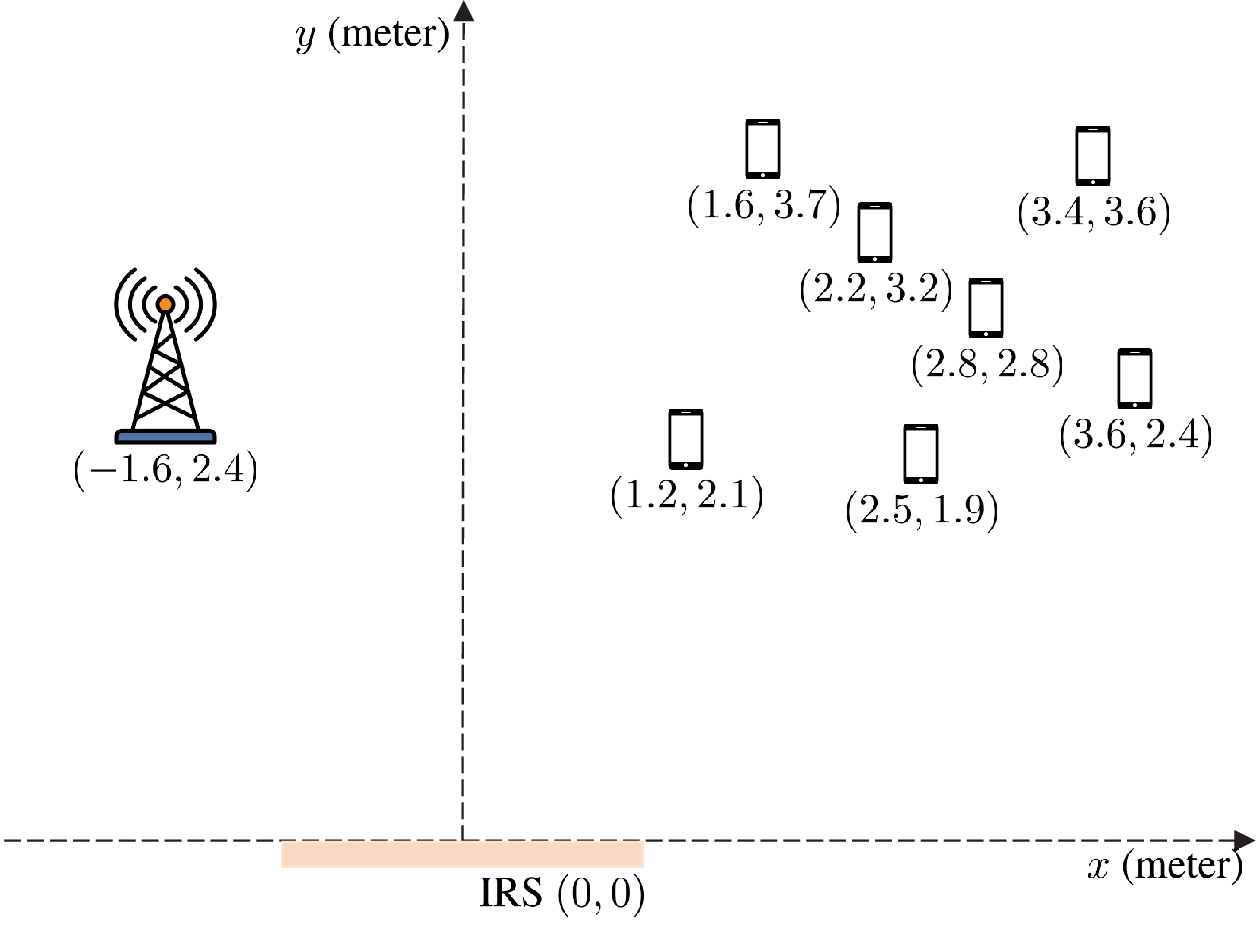}
}
\caption{Our field test at 2.6 GHz in a large-size conference room. Figure (a) is a photo of the test site. Figure (b) is the layout drawing where the position coordinates are in meters.}
\label{Fig field}
\end{figure}

\subsection{Field Tests}
\label{field tests}

We consider at most 7 receiver positions as shown in Fig. \ref{Fig field}. The transmission frequency is 2.6 GHz. The phase shift choice for each RE is restricted to $\{0,\pi\}$. Two different sizes of IRS are tested: one with 588 REs and the other with 294 REs. We consider the low-SNR scenario and the high-SNR scenario by setting the transmit power to be $P=-15$ dBm and $P=-5$ dBm, respectively. The signal bandwidth is 125 KHz. We let the transmitter send signal through quadrature amplitude modulation. The measured SNR is averaged out over approximately 200 received signal samples. We use $T=500$ random samples by default for blind beamforming.

Table \ref{table min SNR field low snr} and Table \ref{table min SNR field high snr} summarize the performance of different methods in the low-SNR scenario and the high-SNR scenario respectively. All IRS-assisted schemes can enhance the minimum SNR in both scenarios compared to the without-IRS case. MV-CSM achieves the highest minimum SNR  in all the 12 cases. The gaps between the minimum SNR achieved by MV-CSM and the best results attained by the rest schemes increase as $N$ increases from 294 to 588 in both the low-SNR and high-SNR scenarios. Specifically, the gaps increase from 2.07 dB, 1.1 dB, and 0.46 dB to 2.66 dB, 1.84 dB, and 1.74 dB for $U=5$, $U=6$, and $U=7$ respectively when $N$ grows in the low-SNR scenario; and increase from 2.15 dB, 1 dB, and 0.7 dB to 3.09 dB, 4.09 dB, and 3.09 dB for $U=5$, $U=6$, and $U=7$ respectively when $N$ grows in the high-SNR scenario. Notice that the position-based method \cite{Ma2022} is always worse than the proposed MV-CSM method because the position-channel model in \cite{Ma2022} deviates a lot from reality even though our test site (which is a large empty room) is already quite simple. The two beam training algorithms (RMS and EMS) achieve lower minimum SNRs than MV-CSM, which is consistent with the discussion in Remark \ref{remark compare RMS}. The underperformance of the channel estimation-based methods (Neural + SDR and Neural + Gradient) also indicates the inaccuracy of channel estimation in our test site. When $N$ is fixed  in some cases, the minimum SNR of zero-phase-shift, without-IRS and the beam training methods does not change as $U$ increases, because the receiver position with the worst SNR performance already exists within the first five receiver positions.

Moreover, we compare the computational efficiencies of the different algorithms in Table \ref{table time field}. It is evident that the two beam training methods (RMS and EMS) have the fastest computational time, while the time of MV-CSM and P-CSM is close to them. In contrast, the three optimization-based algorithms require much larger computational time. 

\begin{table*}[t]
\setlength{\tabcolsep}{8pt}
\centering
\small
\caption{Minimum SNR (in Decibel) in Field Test in the Low-SNR Regime with Transmit Power $P=-15$ dBm}\label{table min SNR field low snr}
\small
\begin{tabular}{lccc|ccc}
\hline
   & \multicolumn{3}{c}{$N=588$} & \multicolumn{3}{c}{$N=294$} \\ \cline{2-7} 
Methods     & $U=5$   & $U=6$   & $U=7$   & $U=5$   & $U=6$   & $U=7$   \\ \hline
MV-CSM    & 8.54  & 7.45 & 6.95   & 7.13  & 6.03  & 5.39    \\ 
P-CSM  &  5.10  & 5.81  & 4.08  & 4.69  & 3.35   & 2.53  \\
Position-Based \cite{Ma2022}  & 4.06   & 3.13  & 2.94   & 2.82    & 2.75  & 2.21  \\
Neural + SDR \cite{Sun2023} & 5.88   & 4.65   & 3.66  & 5.06    & 4.68  & 2.50  \\
Neural + Gradient \cite{Yan2023,Sun2023}  & 4.16 & 3.23  & 2.58  & 4.03    & 4.80   & 1.69  \\
RMS \cite{blind_beamforming_twc}  & 5.21   & 5.21   & 5.21    & 4.93    & 4.94  & 4.96  \\
EMS \cite{An2022}  & 5.13  & 5.13  & 5.13    & 4.62   & 4.66  & 4.64  \\
Zero Phase Shift   & 1.82  & 1.82  & 1.82 & 1.35  & 1.35  & 1.35  \\
Without IRS         & 0.86   & 0.86   & 0.86  & 0.86  & 0.86  & 0.86  \\
\hline
\end{tabular}    
\end{table*}

\begin{table*}[t]
\setlength{\tabcolsep}{8pt}
\centering
\small
\caption{Minimum SNR (in Decibel) in Field Test in the High-SNR Regime with Transmit Power $P=-5$ dBm}\label{table min SNR field high snr}
\small
\begin{tabular}{lccc|ccc}
\hline
   & \multicolumn{3}{c}{$N=588$} & \multicolumn{3}{c}{$N=294$} \\ \cline{2-7} 
Methods     & $U=5$   & $U=6$   & $U=7$   & $U=5$   & $U=6$   & $U=7$   \\ \hline
MV-CSM    & 16.26 & 15.29 & 13.78  & 13.90 & 12.02 & 10.66  \\ 
P-CSM  &  13.14 & 10.60  & 8.59 & 11.62 & 11.02  & 9.94 \\
Position-based \cite{Ma2022}  & 10.02 & 9.74  & 7.22  & 9.06  & 8.46 & 8.06 \\
Neural + SDR \cite{Sun2023} & 13.08 & 11.13 & 10.43  & 10.95  & 10.90 & 9.31 \\
Neural + Gradient \cite{Sun2023,Yan2023}  & 11.65 & 9.93 & 9.01  & 9.51  & 9.12 & 8.35 \\
RMS \cite{blind_beamforming_twc}  & 13.17 & 11.20 & 9.07  & 11.75  & 10.04 & 8.38 \\
EMS \cite{An2022}  & 13.07 & 11.06 & 10.69  & 10.93  & 10.83 & 9.96 \\
Zero Phase Shift   & 9.7 & 9.7  & 9.7  & 7.64  & 6.1 & 6.1 \\
Without IRS         & 6.92 & 6.16 & 5.58 & 6.92 & 6.16 & 5.58 \\
\hline
\end{tabular}    
\end{table*}

\begin{table*}[t]
\centering
\setlength{\tabcolsep}{8pt}
\caption{Running Time (in Seconds) of Different Methods in Field Test}\label{table time field}
\small
\scalebox{1.0}{\begin{tabular}{lccc|ccc}
\hline
   & \multicolumn{3}{c}{$N=588$} & \multicolumn{3}{c}{$N=294$} \\ \cline{2-7} 
Methods     & $U=5$   & $U=6$   & $U=7$   & $U=5$   & $U=6$   & $U=7$   \\ \hline
MV-CSM  & 5.9 & 7 & 10.4 & 2.2 & 2.8 & 3.6    \\ 
P-CSM & 2.6 & 2.8 & 3.1 & 1.2 & 1.4 & 1.8 \\
Position-based \cite{Ma2022}  & 60.2 & 69.1 & 74.5 & 51.3 & 56.1 & 67.4 \\
Neural + SDR \cite{Sun2023} & 84.9 & 98.3 & 114.6 & 46 & 55.2 & 62.6 \\
Neural + Gradient \cite{Sun2023,Yan2023}  & 37.1 & 42.3 & 50.5 & 28.2 & 34.8 & 40.1 \\
RMS \cite{blind_beamforming_twc} & 1.2 & 1.7 & 2.2 & 1.1 & 1.6 & 1.8 \\
EMS \cite{An2022} & 1.3 & 1.7 & 2.1 & 1.1 & 1.5 & 1.9 \\
\hline
\end{tabular} }   
\end{table*}

\subsection{Simulation}

\begin{figure}[t]
\centering
\includegraphics[width=8cm]{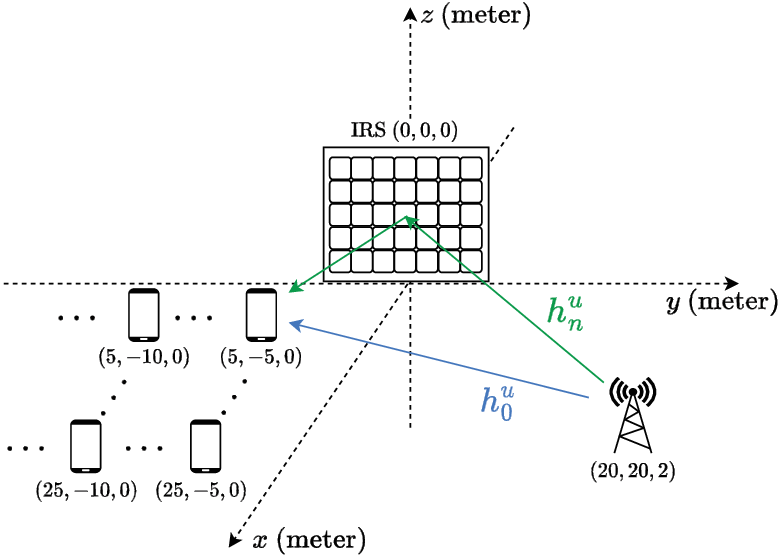}
\caption{Network topology in our simulation. The position coordinates are all in meters. The IRS is placed on the $y$-$z$ plane.}\label{Fig simulation network}
\end{figure}

We further carry out simulations in order to test MV-CSM in more complicated network scenarios. The network topology is shown in Fig. \ref{Fig simulation network}; the $u$-th receiver position is $\big(5\times((u-1)\mod 5+1), -5\times(\lfloor\frac{u-1}{5}\rfloor+1),0\big)$. The number of phase shift candidates is $K=4$, i.e., $\Phi_K=\{0,\frac{\pi}{2},\pi,\frac{3\pi}{2}\}$.   
The transmit power equals $20$ dBm. The background noise power equals $-80$ dBm.

The channel model follows the existing works \cite{Wu2019,Jiang2021a}. The pathloss factors from BS to each position $u$, from BS to IRS, and from IRS to receiver position $u$ are respectively modeled as
\begin{align}
\textrm{PL}_{\textrm{BS},u}&=10^{-(32.6+36.7\log_{10}d_{\textrm{BS},u})/10},\notag\\
\textrm{PL}_{\textrm{BS},\textrm{IRS}}&=10^{-(30+22\log_{10}d_{\textrm{BS},\textrm{IRS}})/10},\notag\\
\textrm{PL}_{\textrm{IRS},u}&=10^{-(30+22\log_{10}d_{\textrm{IRS},u})/10},\notag
\end{align}
where $d_{\textrm{BS},u}$, $d_{\textrm{BS},\textrm{IRS}}$, and $d_{\textrm{IRS},u}$ are the corresponding distances in meters. Under the above settings, the channel coefficients in \eqref{Y:U-IRS} are given by
\begin{align}
&h_{u,0}=\sqrt{\textrm{PL}_{\textrm{BS},u}}\delta_{\textrm{BS},u},\notag\\
&h_{u,n}=\sqrt{\textrm{PL}_{\textrm{BS},\textrm{IRS}}\textrm{PL}_{\textrm{IRS},u}}\delta_{\textrm{BS},n}\delta_{n,u},\notag
\end{align}
where $\delta_{\textrm{BS},u},\delta_{\textrm{BS},\textrm{IRS}},\delta_{\textrm{BS},u},\delta_{u,n}\sim \mathcal{CN}(0,1)$, for each $1\le u\le U$ and each $1\le n\le N$.

\begin{figure}[t]
\centering
\subfigure[]
{
\includegraphics[width=9cm]{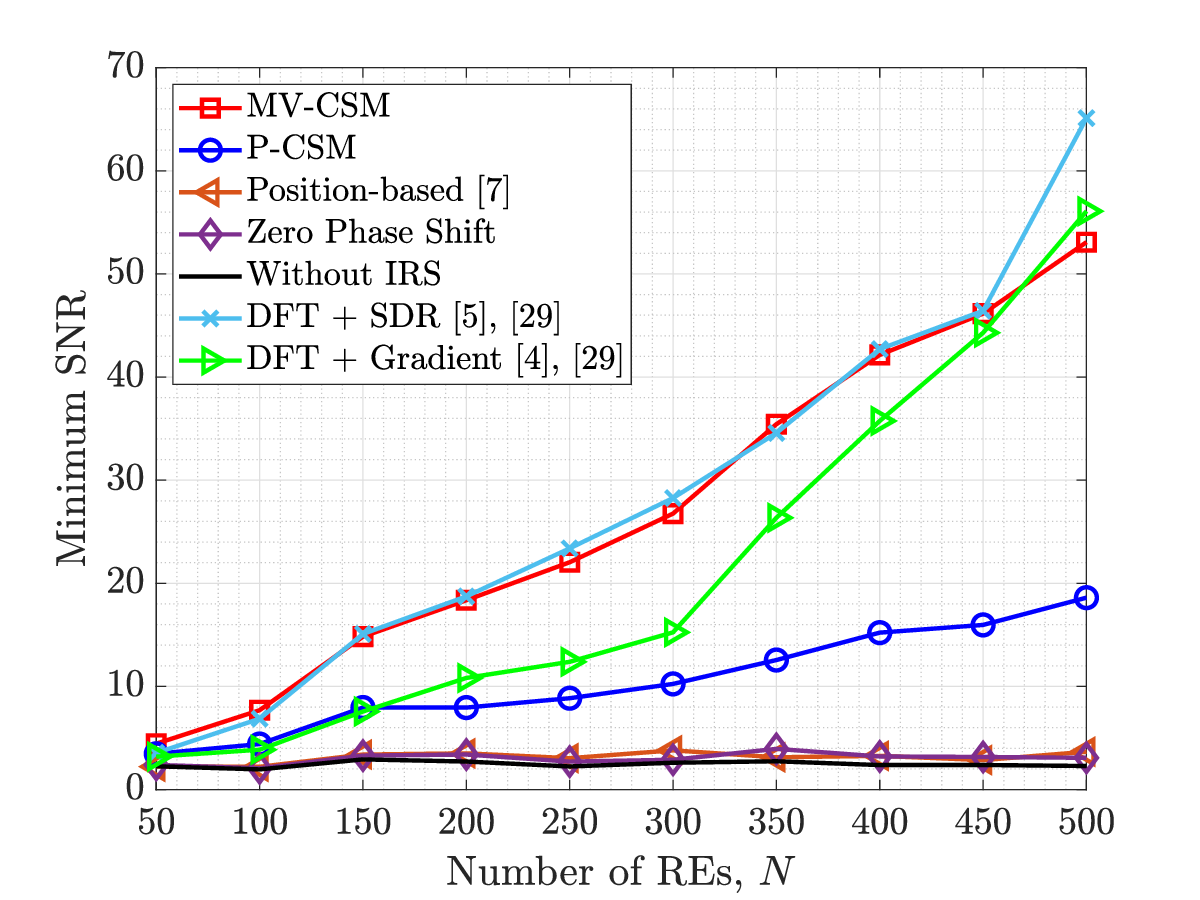}
}\\
\subfigure[]
{
\includegraphics[width=9cm]{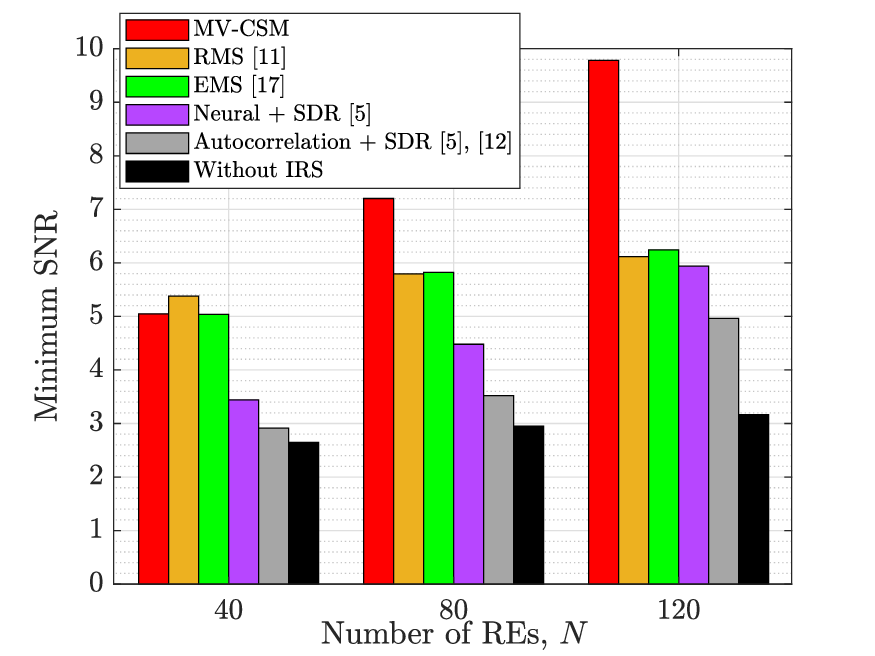}
}
\caption{Minimum SNR versus $N$ when $U=10$. We remark that the minimum SNR is in a linear scale here so as to show the quadratic growth of MV-CSM.}
\label{Fig SNRvsN}
\end{figure}

\begin{figure}[t]
\centering
\includegraphics[width=9cm]{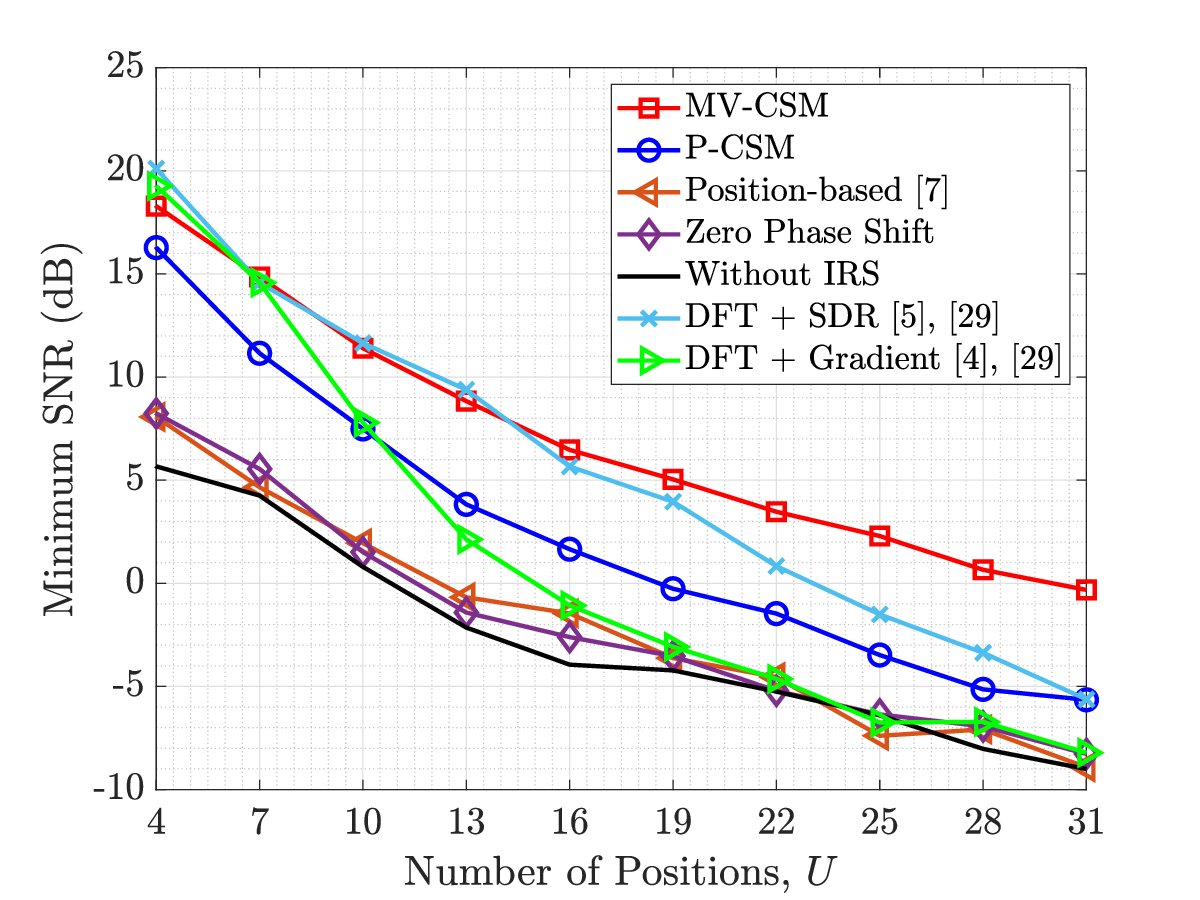}\vspace{-5pt}
\caption{Minimum SNR versus $U$ when $N=200$.}\label{Fig SNRvsU}
\end{figure}

Fig. \ref{Fig SNRvsN} shows the minimum SNR versus the RE number $N$ when there are $U=10$ positions. In Fig. \ref{Fig SNRvsN}(a), the proposed MV-CSM method and the DFT + SDR method achieve the highest minimum SNR when $N\le 450$. MV-CSM is worse than the DFT + SDR method and the DFT + Gradient method when $N=500$. But DFT + SDR and DFT + Gradient require channel estimation. Notice that the growth curve of the minimum SNR achieved by MV-CSM is approximately quadratic in $N$; this result is consistent with Corollary \ref{corollary achievable min}. The position-based method has far worse performance because of the position-channel model error. The DFT + SDR method yields the best performance but at the cost of channel estimation (as discussed shortly).

Fig. \ref{Fig SNRvsN}(b) compares MV-CSM with the beam training methods (RMS and EMS) and the channel estimation-based methods (Neural + SDR and Autocorrelation + SDR) that use the information of received signal power to estimate channels. The proposed MV-CSM method achieves the highest minimum SNR among the considered methods when $N=80$ and $120$, and is close to RMS when $N=40$. It is intuitive to observe that the two beam training methods (RMS and EMS) exhibit similar performance, since both methods achieve their minimum SNR by selecting the sample that attains the maximum minimum SNR. Although channel estimation is applied, the Neural + SDR and Autocorrelation + SDR methods have the worst performance, which is also observed in the field test.

\begin{table}[t]
\caption{Running Time (in Seconds) of Different Methods in Simulation with $U=10$}
\label{table computational time simultion U=10}
\centering
\scalebox{0.85}{\begin{tabular}{lccc}
\hline
\multicolumn{1}{c}{Methods}                                                         & $N=50$  & $N=250$ & $N=500$ \\ \hline
MV-CSM                                                                              & 0.1     & 1.7     & 13.3    \\
P-CSM                                                                               & 0.1     & 1.7     & 13.9    \\
Position-based \cite{Ma2022}                                       & 9.5     & 30.2    & 341.5   \\
DFT + SDR \cite{channel_est_DFT,Sun2023} & 3.4     & 62.1    & 455.3   \\
DFT + Gradient \cite{channel_est_DFT,Yan2023}  & 0.1     & 1.2     & 6.0     \\ \hline
                                                                                    & $N=40$  & $N=80$  & $N=120$ \\ \hline
MV-CSM                                                                              & 0.3     & 0.4     & 0.8     \\
RMS \cite{blind_beamforming_twc}                                                                                & 0.2     & 0.2     & 0.6     \\
EMS      \cite{An2022}                                                                           & 0.2     & 0.3     & 0.5     \\
Neural + SDR \cite{Sun2023}                                            & 102.1   & 107.3   & 114.4   \\
Autocorrelation + SDR \cite{Sun2023,Yan2023a}         & 2884    & 44248   & 282318  \\ \hline
\end{tabular}
}
\end{table}

\begin{table}[t]
\caption{Running Time (in Seconds) of Different Methods in Simulation with $N=200$}
\label{table computational time simultion N=200}
\centering
\scalebox{0.9}{\begin{tabular}{lccc}
\hline
\multicolumn{1}{c}{Methods}                                                         & $U=4$  & $U=16$  & $U=31$  \\ \hline
MV-CSM                                                                              & 1.1    & 0.9     & 1.0     \\
P-CSM                                                                               & 1.2    & 0.9     & 1.0     \\
Position-based \cite{Ma2022}                                       & 23.9   & 20.9    & 23.6    \\
DFT + SDR \cite{channel_est_DFT,Sun2023} & 20.4   & 60.8    & 107.4   \\
DFT + Gradient \cite{channel_est_DFT,Yan2023}  & 0.9    & 0.9     & 1.1     \\ \hline
\end{tabular}
}
\end{table}

Fig. \ref{Fig SNRvsU} shows the minimum SNR versus the position number $U$ when the IRS has $N=200$ REs. MV-CSM attains the highest minimum SNR when $U\ge16$. When $U<16$, the performance of MV-CSM is close to that of the DFT + SDR scheme. The DFT + Gradient method and the DFT + SDR method have the best performance when $U$ is small. When $U$ becomes larger, the DFT + Gradient method gets close to the position-based method; they are both worse than P-CSM, and much worse than MV-CSM. 

Moreover, we compare the running time of different algorithms in Table \ref{table computational time simultion U=10} and Table \ref{table computational time simultion N=200}. Observe that the MV-CSM, P-CSM, DFT + Gradient, RMS and EMS require similar running time. The SDR-based methods  are much more time-consuming in comparison, e.g., the running time of DFT + SDR is more than 30 times higher than that of MV-CSM as shown in Table \ref{table computational time simultion U=10}. Thus, SDR is not suited for the practical configuration of the IRS despite its remarkable min-SNR performance in Fig. \ref{Fig SNRvsN}(a). Among these SDR-based methods, the Autocorrelation + SDR method requires the most computational time as shown in Table \ref{table computational time simultion U=10}, which indicates that estimating the autocorrelation of received signals is not suitable for channel estimation in practice.

\begin{figure}[t]
\centering
\includegraphics[width=9cm]{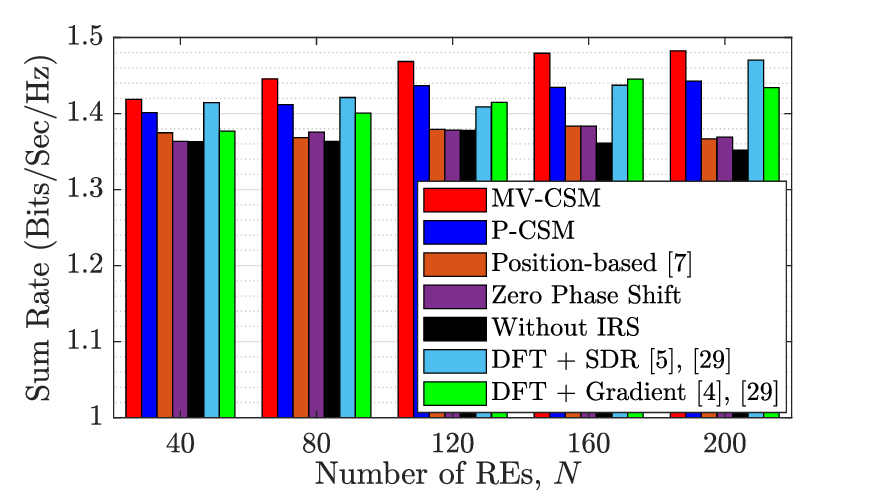}\vspace{-5pt}
\caption{Sum Rate of all positions when $U=10$. We let the base station send a distinct message to each receiver position with uniform power allocation.}\label{Fig sumrate}
\end{figure}

While the above simulations employ the minimum-SNR metric and focus on the scenario where all the receiver positions request the same message, we additionally assess the performance of MV-CSM by utilizing the sum rate as the metric. Specifically, Fig. \ref{Fig sumrate} shows the sum rate of all the receiver positions versus the RE number $N$ when there are $U=10$ positions, in the scenario where the base station sends a distinct message to each position with uniform power allocation. The sum rate of all the considered schemes are close to each other, while MV-CSM still achieves the best results. Notice that the sum rate of all the schemes is close to 1.5 because by using uniform power allocation, the sum rate can be approximated as $R\approx U\log(1+1/(U-1))\approx 1.5$ if we neglect the noise.
The P-CSM method has comparable performance to the two channel estimation-based methods (DFT + SDR and DFT + Gradient), despite the fact that P-CSM does not require channel estimation as MV-CSM. The position-based method continues to exhibit suboptimal performance due to the inaccurate channel modelling.

\section{Conclusion}\label{sec:conclucion}
This work considers coordinating the phase shifts of the different REs of the IRS in order to enhance the SNRs at multiple distributed receiver positions, namely the IRS-assisted coverage enhancement. The proposed blind beamforming approach can be distinguished from the existing methods in two respects: first, blind beamforming does not require any CSI and channel acquisition; second, blind beamforming does not assume any position-channel model. The main idea of blind beamforming is to optimize phase shifts based on the key features of the wireless environment which are extracted from the received signal power data by the conditional sample mean and majority voting. We further provide the asymptotic bounds for both achievability and converse to show that the proposed MV-CSM method guarantees performance close to the global optimum. Our method has been successfully implemented in a prototype system at 2.6 GHz.

\section*{Appendix A: Proof of Theorem \ref{prop achievable}}
Following the proof of \cite{blind_beamforming_twc}, we can show that $\theta_{u,n}=\theta^{\textrm{CPP}}_{u,n}$ w.h.p. for each receiver position $u$ and each RE $n$ so long as $T=\Omega(N^2(\ln NU)^3+N^2U(\ln NU))$, where $\theta_{u,n}$ and $\theta^{\textrm{CPP}}_{u,n}$ are the CSM solution and the CPP solution, respectively, for receiver position $u$. Hence, in the rest of the proof, we just use $\theta^{\textrm{CPP}}_{u,n}$ to denote the CSM solution for position $u$.

In the ideal case where $\theta_{n}^{\textrm{MV-CSM}}=\theta^{\textrm{CPP}}_{u,n}$ for every $u$ and every $n$, the SNR boost due to the IRS can be maximized to be quadratic in $N$ at every position. The central idea of the proof is to show a surprising fact that the ideal case can be achieved asymptotically.

Denote by $\xi_u$ the number of REs with $\theta^{\textrm{MV-CSM}}_{n}=\theta^{\textrm{CPP}}_{u,n}$ for position $u$. We wish to bound $\xi_u$ from below and also from above. Toward this end, introduce an auxiliary variable $p_1$ as 
\begin{align}
p_1=\left\{\!\!
\begin{array}{ll}
(\frac{1}{2})^{U}\times{{U-1}\choose {0.5U-0.5}}
,\!\! & \textrm{if $U$ is odd;}
\vspace{0.5em}\\
(\frac{1}{2})^{U}\times{{U-1}\choose {0.5U-1}}, \!\! & \textrm{if $U$ is even.}
\end{array}
\right.\label{eqn define p1}
\end{align}
By Stirling's formula, we have $p_1\sim\frac{1}{\sqrt{2\pi U}}$ for $U$ sufficiently large. The following lemma gives a lower bound and an upper bound on $\xi_u$, the proof of which is relegated to Appendix D.

\begin{lemma}\label{lemma achievable xi_u}
When the binary random sampling is used, for $U$ and $N$ both sufficiently large with $N=\omega(U^{2})$, it is true that \begin{equation}
\frac{N}{2}+Np_1-\sqrt{N\ln U-N}<\xi_u<\frac{2N}{3}
\end{equation}
for each position $u$ w.h.p. under the MV-CSM method.
\end{lemma}

Notice that the  binary random sampling is critical for Lemma \ref{lemma achievable xi_u}, which ensures that both $\xi_u$ and $N-\xi_u$ (the number of REs with $\theta^{\textrm{MV-CSM}}_{n}\neq\theta^{\textrm{CPP}}_{u,n}$) satisfy Bernoulli distributions. This facilitates the approximation of $\xi_u$ as in Lemma \ref{lemma achievable xi_u} and the following proof. In light of the above lemma, we can now evaluate the SNR for each position $u$. For each $u$, classify the REs into the following two groups:
\begin{align}
\mathcal{N}_{u,1}&=\Big\{n=1,\ldots,N:\theta^{\textrm{MV-CSM}}_{n}=\theta^{\textrm{CPP}}_{u,n}\Big\},\\
\mathcal{N}_{u,2}&=\Big\{n=1,\ldots,N:\theta^{\textrm{MV-CSM}}_{n}\ne\theta^{\textrm{CPP}}_{u,n}\Big\}.
\end{align}
Note that $\xi_u=|\mathcal N_{u,1}|$.

According to Lemma \ref{lemma achievable xi_u}, we have 
\begin{align*}
    &\frac{N}{2}+Np_1-\sqrt{N\ln U-N}< \xi_u<\frac{2N}{3}
\end{align*}
and
\begin{align*}
    &\frac{N}{3}<|\mathcal N_{u,2}|<\frac{N}{2}-Np_1+\sqrt{N\ln U-N}.
\end{align*}
Furthermore, because $N=\omega(U^{2})$ and $p_1\sim \frac{1}{\sqrt{2\pi U}}$, we have $|\mathcal{N}_{u,1}|>|\mathcal{N}_{u,2}|$ and $|\mathcal{N}_{u,1}|-|\mathcal{N}_{u,2}|=\Omega(N/\sqrt{U})$. Besides, it can be shown that $|\mathcal{N}_{u,1}|=\Theta(N)$ and $|\mathcal{N}_{u,2}|=\Theta(N)$. The SNR at position $u$ can now be bounded from below as
\begin{subequations}\label{eqn prove achievable snr}
\begin{align}
&\textrm{SNR}_u \notag\\
&=\frac{P}{\sigma^2}\Bigg|h_{u,0}\!+\!\!\!\sum_{n\in\mathcal{N}_{u,1}}h_{u,n}e^{j\theta^{\textrm{MV-CSM}}_n}+\sum_{n\in\mathcal{N}_{u,2}}h_{u,n}e^{j\theta^{\textrm{MV-CSM}}_n}\Bigg|^2\notag\\
&\ge \frac{P}{\sigma^2}\textrm{Re}\Bigg(|h_{u,0}|+\sum_{n\in\mathcal{N}_{u,1}}c_ue^{j(\angle h_{u,n}+\theta^{\textrm{MV-CSM}}_n-\angle h_{u,0})}\notag\\
&\qquad\quad +\sum_{n\in\mathcal{N}_{u,2}}c_u e^{j(\angle h_{u,n}+\theta^{\textrm{MV-CSM}}_n-\angle h_{u,0})}\Bigg)^2\label{eqn prove achievable snr 1}\\
&=\frac{P}{\sigma^2}\Bigg(|h_{u,0}|+\sum_{n\in\mathcal{N}_{u,1}}c_u\cos(\angle h_{u,n}+\theta^{\textrm{MV-CSM}}_n-\angle h_{u,0})\notag\\
&\qquad +\sum_{n\in\mathcal{N}_{u,2}}c_u\cos(\angle h_{u,n}+\theta^{\textrm{MV-CSM}}_n-\angle h_{u,0})\Bigg)^2,\label{eqn prove achievable snr 2}
\end{align}
\end{subequations}
where \eqref{eqn prove achievable snr 1} is due to the condition $|h_{u,n}|=c_u$ in Assumption \ref{assumption:1}. To further simplify the lower bound in \eqref{eqn prove achievable snr 2}, we need the following lemma:
\begin{lemma}\label{lemma achievable bound M}
Define two new variables to be
\begin{align}
&M_{u,1}=\frac{1}{|\mathcal{N}_{u,1}|}\sum_{n\in\mathcal{N}_{u,1}}c_u\cos(\angle h_{u,n}+\theta^{\textrm{MV-CSM}}_n-\angle h_{u,0})\notag\\
&M_{u,2}=\frac{1}{|\mathcal{N}_{u,2}|}\sum_{n\in\mathcal{N}_{u,2}}c_u\cos(\angle h_{u,n}+\theta^{\textrm{MV-CSM}}_n-\angle h_{u,0}).\notag
\end{align}
For $U$ and $N$ sufficiently large with $N=\omega(U^2)$, it is true that 
\begin{align}
M_{u,1}&\ge \frac{2}{\pi}c_u-\epsilon,\\
M_{u,2}&\ge -\frac{2}{\pi}c_u-\epsilon,
\end{align}
hold w.h.p. for any $\epsilon$ satisfying $\epsilon=\omega(\frac{U^{1/2}}{N^{1/2}})$ and $\epsilon=o(\frac{1}{U^{1/2}})$.
\end{lemma}
The proof of this lemma is relegated to Appendix E.

Plugging the above two inequalities into \eqref{eqn prove achievable snr 2}, we arrive at 
\begin{subequations}\label{eqn achievable sum real}
\begin{align}
&|h_{u,0}|+\sum_{n\in\mathcal{N}_{u,1}}c_u\cos(\angle h_{u,n}+\theta^{\textrm{MV-CSM}}_n-\angle h_{u,0})\notag\\
&\qquad+\sum_{n\in\mathcal{N}_{u,2}}c_u\cos(\angle h_{u,n}+\theta^{\textrm{MV-CSM}}_n-\angle h_{u,0})\notag\\
&\ge \left((|\mathcal{N}_{u,1}|-|\mathcal{N}_{u,2}|)+|\mathcal{N}_{u,2}|\right)\bigg(\frac{2}{\pi}c_u-\epsilon\bigg)\notag\\
&\qquad+ |\mathcal{N}_{u,2}|(-\frac{2}{\pi}c_u-\epsilon)\\
&= \frac{2}{\pi}c_u(|\mathcal{N}_{u,1}|-|\mathcal{N}_{u,2}|)-N\epsilon\label{eqn achievable sum real 1}\\
&= \frac{2}{\pi}c_u\Omega\bigg(\frac{N}{\sqrt{U}}\bigg)-o\bigg(\frac{N}{\sqrt{U}}\bigg)\label{eqn achievable sum real 2}\\
&= \frac{2}{\pi}c_u\Omega\bigg(\frac{N}{\sqrt{U}}\bigg)\label{eqn achievable sum real 3}
\end{align}
\end{subequations}
where \eqref{eqn achievable sum real 2} follows as $\epsilon=o(\frac{1}{U^{1/2}})$ and $|\mathcal{N}_{u,1}|-|\mathcal{N}_{u,2}|=\Omega(\frac{N}{\sqrt{U}})$. Combining \eqref{eqn prove achievable snr} and \eqref{eqn achievable sum real 3} establishes Theorem \ref{prop achievable}. 

\section*{Appendix B: Proof of Theorem \ref{prop converse}}

According to Definition \ref{defiition:good_alg}, a good IRS beamforming algorithm guarantees that
\begin{equation}
    \Bigg|h_{u,0}+\sum_{n=1}^N h_{u,n}e^{j\theta_n}\Bigg|^2\ge |h_{u,0}|^2
\end{equation}
for every $u$, which amounts to
\begin{align}
&\left|\sum_{n=1}^N h_{u,n}e^{j\theta_n}\right|^2+2\textrm{Re}\left(h^*_{u,0}\sum_{n=1}^N h_{u,n}e^{j\theta_n}\right)\ge 0
\end{align}
for every $u$. Thus, for any $b\ge a\ge 1$, we have
\begin{align}
\left|h_{u,0}+\sum_{n=1}^N h_{u,n}e^{j\theta_n}\right|^2\le \left|ah_{u,0}+b\sum_{n=1}^N h_{u,n}e^{j\theta_n}\right|^2
\end{align}
for every $u$. We then bound the sum of SNRs across the $U$ positions as
\begin{align}
&\sum^U_{u=1} \textrm{SNR}_u\notag\\
& \le  \frac{P}{\sigma^2}\sum^U_{u=1}\left|\max_u |h_{u,0}|\cdot e^{j\angle h_{u,0}}+\eta\sum_{n=1}^N c_{\textrm{max}} e^{j\angle h_{u,n}+j\theta_n}\right|^2\notag\\
&=\frac{P}{\sigma^2}\bigg[U\cdot\max_u |h_{u,0}|^2+NU\eta^2 c_{\textrm{max}}^2\notag\\
&\;+\sum_{n=1}^N\sum_u 2\max_u |h_{u,0}|\cdot \eta c_{\textrm{max}} \cos(\theta_n\!+\!\angle h_{u,n}\!-\!\angle h_{u,0})\notag\\
&\; + \sum^N_{n=1}\sum_{n'>n}\sum^U_{u=1} 2\eta^2 c_{\textrm{max}}^2\cos\big(\theta_n\!+\!\angle h_{u,n}-\theta_{n'}-\angle h_{u,n'})\big)\bigg].\label{eqn converse sum snr 1}
\end{align}
If $K$ is raised to its multiple $K'=\alpha K$ where $\alpha>1$ is a positive integer, then the resulting global optimum of $\mathrm{SNR}_\text{min}$ must increase as well since some new choices of phase shifts are considered in addition to the original ones; we remark that $\mathrm{SNR}_\text{min}$ does not necessarily increase if $\alpha>1$ is just a real number. In our case, let $K'=2K\cdot \lceil\frac{U^{\frac{1}{4}}}{2K}\rceil$. Notice that $K'\sim U^{\frac{1}{4}}$ as $U$ becomes large.  We now further bound the terms in \eqref{eqn converse sum snr 1} separately.

For each $n=1,\ldots,N$ and each $k=0,\ldots,K'-1$, we denote by $\mathcal{U}_{k,n}$ the subset of positions which satisfy $\theta_n-\theta^{\textrm{CPP}}_{u,n}={2\pi k}/{K'}$. We can now bound $\sum_u\cos(\theta_n+\angle h_{u,n}-\angle h_{u,0})$ from above as shown in \eqref{eqn converse snr h(k0)h(kn)} at the top of the next page, where \eqref{eqn converse snr h(k0)h(kn) 2} follows since
$$-\frac{\pi}{K'}\le\theta^{\textrm{CPP}}_{u,n}+\angle h_{u,n}-\angle h_{u,0}\le\frac{\pi}{K'},$$
$$\cos\bigg(\frac{2\pi k}{K'}+\theta^{\textrm{CPP}}_{u,n}+\angle h_{u,n}-\angle h_{u,0}\bigg)\le \cos(\frac{2\pi k}{K'}-\frac{\pi}{K'})
,$$
and
\begin{multline*}
    \cos\left(\frac{2\pi (k+\frac{K'}{2}-1)}{K'}+\theta^{\textrm{CPP}}_{u,n}+\angle h_{u,n}-\angle h_{u,0}\right)\\\le-\cos(\frac{2\pi k}{K'}-\frac{\pi}{K'}),
\end{multline*}
for each $k=1,\ldots,K'/2-1$. 

\begin{figure*}
\begin{subequations}\label{eqn converse snr h(k0)h(kn)}
\begin{align}
&\sum^U_{u=1}\cos(\theta_n+\angle h_{u,n}-\angle h_{u,0})\notag\\
&=\sum_{k=0}^{K'-1} \sum_{u\in\mathcal{U}_{k,n}}\cos\bigg(\frac{2\pi k}{K'}+\theta^{\textrm{CPP}}_{u,n}+\angle h_{u,n}-\angle h_{u,0}\bigg)\\
& \le  |\mathcal{U}_{0,n}|+ |\mathcal{U}_{K'-1,n}|+\sum_{k=1}^{ \frac{K'}{2}-1}\sum_{u\in\mathcal{U}_{k,n}}\cos\bigg(\frac{2\pi k}{K'}+\theta^{\textrm{CPP}}_{u,n}+\angle h_{u,n}-\angle h_{u,0}\bigg)\notag\\
&\quad+\sum_{k=1}^{ \frac{K'}{2}-1}\sum_{u\in\mathcal{U}_{k+\frac{K'}{2}-1,n}}\cos\bigg(\frac{2\pi (k+\frac{K'}{2}-1)}{K'}+\theta^{\textrm{CPP}}_{u,n}+\angle h_{u,n}-\angle h_{u,0}\bigg)\label{eqn converse snr h(k0)h(kn) 1}\\
&\le  |\mathcal{U}_{0,n}|+ |\mathcal{U}_{K'-1,n}|+\sum_{k=1}^{\frac{K'}{2}-1}\cos\bigg(\frac{2\pi k}{K'}-\frac{\pi}{K'}\bigg)\big(|\mathcal{U}_{k,n}|-|\mathcal{U}_{k+\frac{K'}{2}-1,n}|\big)\label{eqn converse snr h(k0)h(kn) 2}\\
&\le  2\max_{1\le k\le K'-1} |\mathcal{U}_{k,n}|+\bigg(\max_{1\le k\le K'-1} |\mathcal{U}_{k,n}|-\min_{1\le k\le K'-1} |\mathcal{U}_{k,n}|\bigg)
\sum_{k=1}^{\frac{K'}{2}-1}\left|\cos\bigg(\frac{2\pi k}{K'}-\frac{\pi}{K'}\bigg)\right|\label{eqn converse snr h(k0)h(kn) 3}\\
&\le  2\max_{1\le k\le K'-1} |\mathcal{U}_{k,n}|+
\frac{K'}{2}\bigg(\max_{1\le k\le K'-1} |\mathcal{U}_{k,n}|-\min_{1\le k\le K'-1} |\mathcal{U}_{k,n}|\bigg)\label{eqn converse snr h(k0)h(kn) 4}
\end{align}
\end{subequations}
\hrule
\end{figure*}

Define ${\theta}^{\textrm{CPP}}_{u,n,n'}$ for each $n=1,\ldots,N$, each $n'=n+1,\ldots,N$, and each $u=1,\ldots, U$ to be
\begin{align}
&\theta^\text{CPP}_{u,n,n'}=\arg\min_{\phi\in\Phi_K}\left|\angle\left(\frac{h_{u,n}e^{j\phi}}{h_{u,n'}}\right)\right|.
\end{align}
Notice that the value of ${\theta}^{\textrm{CPP}}_{u,n,n'}$ is uniformly distributed on the discrete set $\big\{0,\frac{2\pi}{K'},\ldots,\frac{2\pi(K'-1)}{K'}\big\}$. For each $k=0,\ldots,K'-1$, we denote by $\mathcal{U}_{k,n,n'}$ the subset of positions which satisfy $\theta_n-\theta_{n'}-{\theta}^{\textrm{CPP}}_{u,n,n'}=\frac{2\pi k}{K'}$. Repeating the procedure in \eqref{eqn converse snr h(k0)h(kn)}, we obtain
\begin{align}
&\sum^U_{u=1} \cos(\theta_n-\theta_{n'}+\angle h_{u,n}-\angle h_{u,n'})\notag\\
&= \sum_{k=0}^{K'-1}\sum_{u\in \mathcal{U}_{k,n,n'}}\cos\bigg(\frac{2\pi k}{K'}+{\theta}^{\textrm{CPP}}_{u,n,n'}+\angle h_{u,n}-\angle h_{u,n'}\bigg)\notag\\
&\le  2\cdot\max_{1\le k\le K'-1} |\mathcal{U}_{k,n,n'}|\notag\\
&\quad +\frac{K'}{2}\Big(\max_{1\le k\le K'-1} |\mathcal{U}_{k,n,n'}|-\min_{1\le k\le K'-1}|\mathcal{U}_{k,n,n'}|\Big)\label{eqn converse snr h(kn)h(kn')}
\end{align}


We now introduce a lemma that helps further bound the terms $\max_k |\mathcal{U}_{k,n}|$, $\max_k |\mathcal{U}_{k,n,n'}|$, $\min_k |\mathcal{U}_{k,n}|$ and $\min_k |\mathcal{U}_{k,n,n'}|$ in \eqref{eqn converse snr h(k0)h(kn) 4} and \eqref{eqn converse snr h(kn)h(kn')}, the proof of which is stated in Appendix F.
\begin{lemma}\label{lemma converse bound U}
Recall that $K'=2K\cdot \lceil\frac{U^{\frac{1}{4}}}{2K}\rceil$.
For $N$ and $U$ sufficiently large, we have 
\begin{align*}
\max_k |\mathcal{U}_{k,n}|&<\frac{U}{K'}+\sqrt{U(\ln N^2K')},\\
\max_k |\mathcal{U}_{k,n,n'}|&<\frac{U}{K'}+\sqrt{U(\ln N^2K')},\\
\min_k |\mathcal{U}_{k,n}|&>\frac{U}{K'}-\sqrt{U(\ln N^2K')},\\
\min_k |\mathcal{U}_{k,n,n'}|&>\frac{U}{K'}-\sqrt{U(\ln N^2K')}
\end{align*}
hold true w.h.p. 
for each $k=0,\ldots,K'-1$, each $n=1,\ldots,N$, and each $n'\ne n$.
\end{lemma}

Substituting \eqref{eqn converse snr h(k0)h(kn)}, \eqref{eqn converse snr h(kn)h(kn')}, and the bounds of Lemma \ref{lemma converse bound U} into \eqref{eqn converse sum snr 1}, we have w.h.p. that
\begin{subequations}\label{eqn converse sum SNR final}
\begin{align}
&\sum^U_{u=1} \textrm{SNR}_u\notag\\
&\le \frac{P}{\sigma^2}\Bigg[U\max_u |h_{u,0}|^2\!+\!NU\eta^2 c_{\textrm{max}}^2\notag\\
&\quad +\left(2\max_u |h_{u,0}|\cdot \eta c_{\textrm{max}}N+2\eta^2 c_{\textrm{max}}^2\frac{N^2-N}{2}\right)\cdot\notag\\
&\quad \left(2\left(\frac{U}{K'}+\sqrt{U(\ln N^2K')}\right)+K'\sqrt{U(\ln N^2K')}\right)\Bigg]\notag\\
&=\frac{P}{\sigma^2}\left[\eta^2 c_{\textrm{max}}^2\Theta(NU)+\eta^2 c_{\textrm{max}}^2\Theta(N^2)\Theta(U^{3/4}\sqrt{\ln NU})\right].\label{eqn converse sum SNR final 1}
\end{align}
\end{subequations}
Combining the above bound with the fact that $U\times\mathrm{SNR}_{\min}\le \sum^N_{u=1}\mathrm{SNR}_u$, 
we verify Theorem \ref{prop converse}.

\section*{Appendix C: Proof of Theorem \ref{prop achievable P-CSM}}

The main idea of the present proof is similar to that of the proof of Theorem \ref{prop achievable}, i.e., we show that every position $u$ can enjoy the ideal situation $\theta_{n}^{\textrm{P-CSM}}=\theta^{\textrm{CPP}}_{u,n}$ for almost every RE $n$ when P-CSM is applied. 

Denote by $\mathcal N_u$ the subset of REs assigned to position $u$ for P-CSM.
With respect to each position $u$, we still classify the REs of $\mathcal N_u$ into two groups: 
\begin{align}
\mathcal{N}_{u,1}&=\Big\{n=1,\ldots,N:\theta^{\textrm{P-CSM}}_{n}=\theta^{\textrm{CPP}}_{u,n}\Big\},\\
\mathcal{N}_{u,2}&=\Big\{n=1,\ldots,N:\theta^{\textrm{P-CSM}}_{n}\ne\theta^{\textrm{CPP}}_{u,n}\Big\}.
\end{align}
Then we bound the deviations of the sizes of these two groups in the following lemma whose proof is stated in Appendix G.
\begin{lemma}\label{lemma achievable PCSM N size}
For $U$ and $N$ both sufficiently large, it is true w.h.p. that
\begin{align*}
\bigg|\mathcal{N}_{u,1}-\frac{N(U+1)}{2U}\bigg|&<\sqrt{N\ln U},\\
\bigg|\mathcal{N}_{u,2}-\frac{N(U-1)}{2U}\bigg|&<\sqrt{N\ln U},
\end{align*}
for each position $u$ under the P-CSM method.
\end{lemma}

Since we assume $N=\omega(U^3)$ in Theorem \ref{prop achievable P-CSM}, it immediately follows from
Lemma \ref{lemma achievable PCSM N size} that $|\mathcal{N}_{u,1}|-|\mathcal{N}_{u,2}|=\Theta(\frac{N}{U})$. Similar to the MV-CSM case of \eqref{eqn prove achievable snr}, the SNR by P-CSM at each position $u$ can be bounded from below as
\begin{align}
&\textrm{SNR}_u\notag\\
&\ge \frac{P}{\sigma^2}\bigg(|h_{u,0}|+\sum_{n\in\mathcal{N}_{u,1}}c_u\cos(\angle h_{u,n}+\theta^{\textrm{P-CSM}}_n-\angle h_{u,0})\notag\\
&\qquad +\sum_{n\in\mathcal{N}_{u,2}}c_u\cos(\angle h_{u,n}+\theta^{\textrm{P-CSM}}_n-\angle h_{u,0})\bigg)^2.\label{eqn prove achievable snr PCSM}
\end{align}
Moreover, letting
\begin{align}
M_{u,1}&=\frac{1}{|\mathcal{N}_{u,1}|}\sum_{n\in\mathcal{N}_{u,1}}c_u\cos(\angle h_{u,n}+\theta^{\textrm{P-CSM}}_n-\angle h_{u,0}),\notag\\
M_{u,2}&=\frac{1}{|\mathcal{N}_{u,2}|}\sum_{n\in\mathcal{N}_{u,2}}c_u\cos(\angle h_{u,n}+\theta^{\textrm{P-CSM}}_n-\angle h_{u,0}),\notag
\end{align} 
the following lemma bounds their values from below.
\begin{lemma}\label{lemma achievable bound M PCSM}
For $U$ and $N$ both sufficiently large with $N=\omega(U^3)$, we have 
\begin{align}
M_{u,1}&\ge \frac{2}{\pi}c_u-\varepsilon\\
M_{u,2}&\ge -\frac{2}{\pi}c_u-\varepsilon
\end{align}
hold true w.h.p. for each position $u$, where $\varepsilon$ is an arbitrary variable satisfying $\varepsilon=\omega(\frac{U^{1/2}}{N^{1/2}})$ and $\varepsilon=o(\frac{1}{U})$.
\end{lemma}

The above lemma is verified in Appendix H. Equipped with Lemma \ref{lemma achievable bound M PCSM}, we can bound the term on the right-hand side of \eqref{eqn prove achievable snr PCSM} from below as
\begin{subequations}\label{eqn achievable sum real PCSM}
\begin{align}
&|h_{u,0}|+\sum_{n\in\mathcal{N}_{u,1}}c_u\cos(\angle h_{u,n}+\theta^{\textrm{P-CSM}}_n-\angle h_{u,0})\notag\\
&\quad+\sum_{n\in\mathcal{N}_{u,2}}c_u\cos(\angle h_{u,n}+\theta^{\textrm{P-CSM}}_n-\angle h_{u,0})\notag\\
&\ge|\mathcal{N}_{u,1}|\bigg(\frac{2}{\pi}c_u-\varepsilon\bigg)- |\mathcal{N}_{u,2}|\Big(\frac{2}{\pi}c_u+\varepsilon\Big)\\
&=\frac{2}{\pi}c_u(|\mathcal{N}_{u,1}|\!-\!|\mathcal{N}_{u,2}|)-N\varepsilon\label{eqn achievable sum real 1 PCSM}\\
&= \frac{2}{\pi}c_u\Theta\bigg(\frac{N}{U}\bigg)-o\bigg(\frac{N}{U}\bigg)\label{eqn achievable sum real 2 PCSM}\\
&= \frac{2}{\pi}c_u\Theta\bigg(\frac{N}{U}\bigg),\label{eqn achievable sum real 3 PCSM}
\end{align}
\end{subequations}
where \eqref{eqn achievable sum real 2 PCSM} follows since $\varepsilon=o(\frac{1}{U})$ and $|\mathcal{N}_{u,1}|-|\mathcal{N}_{u,2}|=\Theta(\frac{N}{U})$. Substituting the above lower bound in \eqref{eqn prove achievable snr PCSM} establishes Theorem \ref{prop achievable P-CSM}. 

\section*{Appendix D: Proof of Lemma \ref{lemma achievable xi_u}}

We first evaluate $\mathbb{P}(\theta^{\textrm{MV-CSM}}_{n}=\theta^{\textrm{CPP}}_{u,n})$ for a particular RE $n$ and a particular position $u$. Since the channel phases are uniformly distributed according to Assumption \ref{assumption:1}, each $\theta^{\textrm{CPP}}_{u,n}$ is uniformly distributed on $\{0,\pi\}$. By the binomial distribution, $\mathbb{P}\{\theta^{\textrm{MV-CSM}}_{n}=\theta^{\textrm{CPP}}_{u,n}\}$ is given by
\begin{align}
&\mathbb{P}\{\theta^{\textrm{MV-CSM}}_{n}=\theta^{\textrm{CPP}}_{u,n}\}\notag\\
&=\left\{\!\!
\begin{array}{ll}
\sum_{s=\frac{U-1}{2}}^{U-1}\binom{U-1}{s}(\frac{1}{2})^{U-1},\!\! & \textrm{if $U$ is odd}
\vspace{0.5em}\\
\sum_{s=\frac{U}{2}}^{U-1}\binom{U-1}{s}(\frac{1}{2})^{U-1}\!+\!\frac{1}{2}\binom{U-1}{\frac{U}{2}-1}(\frac{1}{2})^{U-1}, \!\! & \textrm{if $U$ is even}
\end{array}
\right.\notag\\
&=\frac{1}{2}+p_1,\notag
\end{align}
where $p_1$ is defined in \eqref{eqn define p1}. Thus, $\xi_u$ has a binomial distribution $\mathcal B(N,\frac{1}{2}+p_1)$. Letting $\bar{\xi}=\frac{N}{2}+Np_1-\sqrt{N\ln U-N}$, by Hoeffding's inequality \cite{Hoeffding}, we obtain
\begin{align}
\mathbb{P}\left(\xi_u \le \bar{\xi}\right)&<\mathbb{P}\bigg\{\left|\xi_u-\frac{N}{2}-Np_1\right|\ge \sqrt{N\ln U-N} \bigg\}\notag\\
&\le  2e^{-\frac{2(N\ln U-N)}{N}} = \frac{2e^2}{U^2}\label{eqn barxi prob}
\end{align}
Similarly, we have
\begin{align}
\mathbb{P}\bigg\{\xi_u \ge \frac{2N}{3}\bigg\}&=\mathbb{P}\bigg\{\xi_u-\frac{N}{2}-Np_1\ge \frac{N}{6}-Np_1\bigg\}\notag\\
& \le  e^{-\frac{2(\frac{N}{6}-Np_1)^2}{N}}\notag\\
&= e^{-\Theta(N)}\label{eqn 2N/3 prob}
\end{align}
Combining \eqref{eqn barxi prob} and \eqref{eqn 2N/3 prob} yields
\begin{align}
&\mathbb{P}\bigg\{\bar{\xi}< \xi_u<\frac{2N}{3},\;\forall u\bigg\}\notag\\
& =1-\mathbb{P}\bigg\{\xi_u\le \bar{\xi} \;\;\text{or}\;\; \xi_u\ge\frac{2N}{3},\;\exists u\bigg\}\notag\\
&\ge 1-\sum_u\mathbb{P}\{\xi_u\le \bar{\xi}\}-\sum_u\mathbb{P}\bigg\{\xi_u\ge \frac{2N}{3}\bigg\}\notag\\
& \ge  1-\frac{2e^2}{U}-Ue^{-\Theta(N)},\notag
\end{align}
which approaches 1 as $U$ and $N$ are sufficiently large and $N=\omega(U^{2})$.

\section*{Appendix E: Proof of Lemma \ref{lemma achievable bound M}}
For any $n\in \mathcal{N}_{u,1}$, we have $\theta^{\textrm{MV-CSM}}_n=\theta^{\textrm{CPP}}_{u,n}$ and hence $(\angle h_{u,n}+\theta^{\textrm{MV-CSM}}_n-\angle h_{u,0})\in[-\pi/2,\pi/2]$ according to the CPP method in \eqref{CPP}. Since the channel phases are uniformly distributed as in Assumption \ref{assumption:1}, a crucial observation can be made that $(\angle h_{u,n}+\theta^{\textrm{MV-CSM}}_n-\angle h_{u,0})$ has a uniform distribution $ \mathcal U[-\pi/2,\pi/2]$ whenever $n\in \mathcal{N}_{u,1}$. Likewise, $(\angle h_{u,n}+\theta^{\textrm{MV-CSM}}_n-\angle h_{u,0})$ has a uniform distribution $\mathcal U[\pi/2,3\pi/2]$ whenever $n\in \mathcal{N}_{u,2}$. Consequently, we have
\begin{align}
&\mathbb{E}[c_u\cos(\angle h_{u,n}+\theta^{\textrm{MV-CSM}}_n-\angle h_{u,0})]=\left\{
\begin{array}{ll}\!\!\!
\frac{2c_u}{\pi} &\!\!\!\!\! \textrm{if }  n\in \mathcal{N}_{u,1};\\
\!\!\!-\frac{2c_u}{\pi} &\!\!\!\!\! \textrm{if } n\in \mathcal{N}_{u,2}.
\end{array}
\right. \!\!\!\notag
\end{align}
Since $N=\omega(U^{2})$, there must exist an auxiliary variable $\epsilon$ that satisfies $\epsilon=\omega(\frac{U^{1/2}}{N^{1/2}})$ and $\epsilon=o(\frac{1}{U^{1/2}})$. Further, by Chebyshev's inequality, we have
\begin{subequations}\label{eqn chebyshev}
\begin{align}
\mathbb{P}\left\{\left|M_{u,1}-\frac{2c_u}{\pi}\right|\ge \epsilon\right\}&\le  \frac{\bar{\sigma}_u^2}{|\mathcal{N}_{u,1}|\epsilon^2}\label{eqn chebyshev 1},\\
\mathbb{P}\left\{\left|M_{u,2}+\frac{2c_u}{\pi}\right|\ge \epsilon\right\}&\le  \frac{\bar{\sigma}_u^2}{|\mathcal{N}_{u,2}|\epsilon^2}\label{eqn chebyshev 3},
\end{align}
\end{subequations}
where $\bar{\sigma}_u^2$ is the variance of $c_u\cos(\angle h_{u,n}+\theta^{\textrm{MV-CSM}}_n-\angle h_{u,0})$ for each $n$. Using the union bound, we further derive 
\begin{subequations}\label{eqn M prob}
\begin{align}
& \mathbb{P}\left\{\left|M_{u,1}-\frac{2c_u}{\pi}\right|\ge \epsilon \textrm{ or } \left|M_{u,2}+\frac{2c_u}{\pi}\right|\ge \epsilon,\exists u\right\}\notag\\
& \le  \sum^U_{u=1}\left[\mathbb{P}\left\{\left|M_{u,1}-(\frac{2}{\pi}c_u)\right|\ge \epsilon\right\}\right.\notag\\
&\qquad\qquad+\left.\mathbb{P}\left\{\left|M_{u,2}\!-\!(-\frac{2}{\pi}c_u)\right|\ge\epsilon\right\}\right]\\
& \le  \sum^U_{u=1} \frac{\bar{\sigma}_u^2}{\epsilon^2}\left[\frac{1}{|\mathcal{N}_{u,1}|}+\frac{1}{|\mathcal{N}_{u,2}|}\right]\label{eqn M prob 1}\\
&= \frac{\sum^U_{u=1} \bar{\sigma}_u^2}{U}\frac{U}{\epsilon^2}\frac{2}{\Theta(N)}\label{eqn M prob 2}\\
&=  \frac{\sum^U_{u=1} \bar{\sigma}_u^2}{U}\frac{U}{\omega(U/N)}\frac{2}{\Theta(N)},\label{eqn M prob 3}
\end{align}
\end{subequations}
where \eqref{eqn M prob 1} follows by \eqref{eqn chebyshev} while \eqref{eqn M prob 2} follows as $|\mathcal{N}_{u,1}|=|\mathcal{N}_{u,2}|=\Theta(N)$. Clearly, \eqref{eqn M prob 3} approaches 0 for $U$ and $N$ sufficiently large, so Lemma \ref{lemma achievable bound M} can be readily verified.

\section*{Appendix F: Proof of Lemma \ref{lemma converse bound U}}
For each $k\in\{0,\ldots,K'-1\}$, write $\mathcal{U}'_{k,n}= \big\{u:\theta^{\textrm{CPP}}_{u,n}=\frac{2\pi k}{K'}\big\}$ and $\mathcal{U}'_{k,n,n'}= \big\{u:{\theta}^{\textrm{CPP}}_{u,n,n'}=\frac{2\pi k}{K'}\big\}$. It is evident that
\begin{align}
\max_k |\mathcal{U}'_{k,n}|&=\max_k |\mathcal{U}_{k,n}|,\notag\\
\max_k|\mathcal{U}'_{k,n,n'}|&=\max_k |\mathcal{U}_{k,n,n'}|,\notag\\
\min_k |\mathcal{U}'_{k,n}|&=\!\min_k|\mathcal{U}_{k,n}|,\notag\\
\min_k |\mathcal{U}'_{k,n,n'}|&=\min_k |\mathcal{U}_{k,n,n'}|\label{eqn upper bound max=max a}.
\end{align}
In the rest of the proof, we use the variables on the left-hand side of \eqref{eqn upper bound max=max a} to replace the variables on the right-hand side.

Because $(\angle h_{u,n}-\angle h_{u,0})$ is uniformly distributed on $[0,2\pi)$, $\theta^{\textrm{CPP}}_{u,n}$ is uniformly distributed on $\big\{0,\frac{2\pi }{K'},\ldots,\frac{2\pi (K'-1)}{K'}\big\}$, so $|\mathcal{U}'_{k,n}|$ has a binomial distribution $\mathcal B(U,\frac{1}{K'})$. Then Hoeffding's inequality leads us to
\begin{align}
&\mathbb{P}\bigg\{\left|\big|\mathcal{U}'_{k,n}\big|-\frac{U}{K'}\right|\ge\sqrt{U(\ln N^2K')} \bigg\}\notag\\
&\le 2e^{-\frac{2U\ln(N^2K')}{U}}\notag\\
&=\frac{2}{N^4(K')^2}.\label{eqn converse bound U1}
\end{align}
Likewise, for any tuple $(k,n,n')$ with $n'\neq n$, because $\theta^{\textrm{CPP}}_{u,n,n'}$ is uniformly distributed on $\{0,\frac{2\pi }{K'},\ldots,\frac{2\pi (K'-1)}{K'}\}$, we have $|\mathcal{U}'_{k,n,n'}|$ distributed as $\mathcal B(U,\frac{1}{K'})$. Again, by Hoeffding's inequality, we arrive at 
\begin{align}
\mathbb{P}\left\{\left|\big|\mathcal{U}'_{k,n,n'}\big|-\frac{U}{K'}\right|\ge \sqrt{U(\ln N^2K')} \right\}\le\frac{2}{N^4(K')^2}.\label{eqn converse bound U2}
\end{align}
Now, denote by $\mathcal{E}$ the event that there exists a pair $(k,n)$ such that $\left||\mathcal{U}'_{k,n}|-\frac{U}{K'}\right|\ge \sqrt{U(\ln N^2K')}$, and denote by $\mathcal{F}$ the event that there exists a tuple $(k,n,n')$ with $n\neq n'$ such that $\left||\mathcal{U}'_{k,n,n'}|-\frac{U}{K'}\right|\ge \sqrt{U(\ln N^2K')}$. By virtue of the union bound, we show that
\begin{align}
&\mathbb{P}(\mathcal{E}\cup \mathcal{F})\notag\\
& \le  \sum_{k,n} \mathbb{P}\left\{\left|\big|\mathcal{U}'_{k,n}\big|-\frac{U}{K'}\right|\ge \sqrt{U(\ln N^2K')} \right\}\notag\\
&\quad +\sum_{k,n}\sum_{n'\neq n'} \mathbb{P}\left\{\left|\big|\mathcal{U}'_{k,n,n'}\big|-\frac{U}{K'}\right|\!\ge\! \sqrt{U(\ln N^2K')} \right\}\notag\\
& \le  \frac{2}{N^2K'},
\end{align}
which bound tends to 0 for $N$ and $U$ sufficiently large. Hence, we have w.h.p that $\left|\big|\mathcal{U}'_{k,n}\big|-\frac{U}{K'}\right|< \sqrt{U(\ln N^2K')}$  for every $(k,n)$, and also have w.h.p. that $\left|\big|\mathcal{U}'_{k,n,n'}\big|-\frac{U}{K'}\right|< \sqrt{U(\ln N^2K')}$ for every $(k,n,n')$ with $n\neq n'$. After we recover the variables on the right-hand side of \eqref{eqn upper bound max=max a}, the proof is completed.

\section*{Appendix G: Proof of Lemma \ref{lemma achievable PCSM N size}}
The proof here follows that of Lemma \ref{lemma achievable xi_u} in Appendix D closely. Consider a particular position $u$. Recall that $\mathcal N_u$ is the subset of REs assigned to position $u$ for P-CSM. Define $\mathcal{N}_{u}^{\textrm{rest}}= \{n:n\in\mathcal{N}_{u'} \textrm{ with } u'\neq u\}$. 
We further partition $\mathcal{N}_{u}^{\textrm{rest}}$ into $\mathcal{N}'_{u,1}$ and $\mathcal{N}'_{u,2}$, the former with $\theta^{\textrm{P-CSM}}_{n}=\theta^{\textrm{CPP}}_{u,n}$ and the latter with $\theta^{\textrm{P-CSM}}_{n}\neq\theta^{\textrm{CPP}}_{u,n}$. Since $\theta^{\textrm{CPP}}_{u,n}$ is uniformly distributed on $\{0,\pi\}$ as in Assumption \ref{assumption:1} holds, we have
\begin{align}
\mathbb{P}\big\{\theta^{\textrm{P-CSM}}_{n}=\theta^{\textrm{CPP}}_{u,n}\big\}=\mathbb{P}\big\{\theta^{\textrm{P-CSM}}_{n}\neq\theta^{\textrm{CPP}}_{u,n}\big\}=\frac{1}{2}\notag
\end{align} 
for every $n\in \mathcal{N}_{u}^{\textrm{rest}}$. As a result, $|\mathcal{N}'_{u,1}|$ has a binomial distribution $\mathcal B(|\mathcal{N}_{u}^{\textrm{rest}}|,\frac{1}{2})$. By Hoeffding's inequality, we have
\begin{align}
\mathbb{P}\bigg\{\bigg|\big|\mathcal{N}'_{u,1}\big|-\frac{|\mathcal{N}_{u}^{\textrm{rest}}|}{2}\bigg|\ge \sqrt{N\ln U} \bigg\}&\le 2e^{-\frac{2(N\ln U)}{|\mathcal{N}_{u}^{\textrm{rest}}|}}\notag\\
&<2e^{-2\ln U}.\label{eqn barxi prob PCSM}
\end{align}
We then use the union bound to obtain
\begin{align}
&\mathbb{P}\left\{\left||\mathcal{N}'_{u,1}|-\frac{|\mathcal{N}_{u}^{\textrm{rest}}|}{2}\right|\ge \sqrt{N\ln U} ,\exists u\right\}\notag\\
&\le \sum^U_{u=1} \mathbb{P}\left(\left||\mathcal{N}'_{u,1}|-\frac{|\mathcal{N}_{u}^{\textrm{rest}}|}{2}\right|\ge \sqrt{N\ln U} \right)\notag\\
& \le  2Ue^{-2\ln U},\label{eqn PCSM N size}
\end{align}
which bound approaches 0 for $U$ sufficiently large. Combining \eqref{eqn PCSM N size} with the identity $|\mathcal{N}'_{u,2}|=|\mathcal{N}_{u}^{\textrm{rest}}|-|\mathcal{N}'_{u,1}|$ shows that
\begin{align}
\bigg|\mathcal{N}'_{u,1}-\frac{|\mathcal{N}_{u}^{\textrm{rest}}|}{2}\bigg|&<\sqrt{N\ln U},\\
\bigg|\mathcal{N}'_{u,2}-\frac{|\mathcal{N}_{u}^{\textrm{rest}}|}{2}\bigg|&<\sqrt{N\ln U},
\end{align}
hold true w.h.p. for every position $u$. Finally, since $|\mathcal{N}_{u,1}|=|\mathcal{N}'_{u,1}|+N-|\mathcal{N}_{u}^{\textrm{rest}}|$, $|\mathcal{N}_{u,2}|=|\mathcal{N}'_{u,2}|$, and $|\mathcal{N}_{u}^{\textrm{rest}}|\sim\frac{N(U-1)}{U}$, the result of Lemma \ref{lemma achievable PCSM N size} can be verified.


\section*{Appendix H: Proof of Lemma \ref{lemma achievable bound M PCSM}}
The proof here follows that of Lemma \ref{lemma achievable bound M} in Appendix E closely. Because $(\angle h_{u,n}+\theta^{\textrm{P-CSM}}_n-\angle h_{u,0})$ has a uniform distribution $\mathcal U[-\pi/2,\pi/2]$ for each $n\in \mathcal{N}_{u,1}$, and $(\angle h_{u,n}+\theta^{\textrm{P-CSM}}_n-\angle h_{u,0})$ has a uniform distribution $ U[\pi/2,3\pi/2]$ for each $n\in \mathcal{N}_{u,2}$, we have
\begin{align}
&\mathbb{E}\big[\cos(\angle h_{u,n}+\theta^{\textrm{P-CSM}}_n-\angle h_{u,0})\big]=\left\{
\begin{array}{ll}\!\!
\frac{2}{\pi}, & \textrm{if }  n\in \mathcal{N}_{u,1};\\
-\frac{2}{\pi}, & \textrm{if } n\in \mathcal{N}_{u,2}.
\end{array}
\right.\notag
\end{align}
Because of the assumption $N=\omega(U^{3})$, there always exists an auxiliary variable $\varepsilon$ satisfying $\varepsilon=\omega(\frac{U^{1/2}}{N^{1/2}})$ and $\varepsilon=o(\frac{1}{U})$. Applying Chebyshev's inequality gives
\begin{subequations}\label{eqn chebyshev PCSM}
\begin{align}
\mathbb{P}\left\{\left|M_{u,1}-\frac{2}{\pi}c_u\right|\ge \varepsilon\right\}&\le  \frac{\bar{\sigma}_u^2}{|\mathcal{N}_{u,1}|\epsilon^2},\label{eqn chebyshev PCSM 1}\\
\mathbb{P}\left\{\left|M_{u,2}+\frac{2}{\pi}c_u\right|\ge \varepsilon\right\}&\le  \frac{\bar{\sigma}_u^2}{|\mathcal{N}_{u,2}|\epsilon^2},\label{eqn chebyshev PCSM 3}
\end{align}
\end{subequations}
where $\bar{\sigma}_u^2$ is the variance of $c_u\cos(\angle h_{u,n}+\theta^{\textrm{P-CSM}}_n-\angle h_{u,0})$; note that the value of $\bar{\sigma}_u^2$ is independent of $n$. Again, in light of the union bound, we show that
\begin{subequations}\label{eqn M prob PCSM}
\begin{align}
& \mathbb{P}\left\{\left|M_{u,1}-\frac{2}{\pi}c_u\right|\ge \varepsilon \textrm{ or } \left|M_{u,2}+\frac{2}{\pi}c_u\right|\ge \varepsilon,\exists u\right\}\notag\\
&\le \sum^U_{u=1}\mathbb{P}\left\{\left|M_{u,1}-(\frac{2}{\pi}c_u)\right|\ge \varepsilon\right\}\notag\\
&\qquad+\sum^U_{u=1}\mathbb{P}\left\{\left|M_{u,2}-(-\frac{2}{\pi}c_u)\right|\ge \varepsilon\right\}\\
&\le  \sum^U_{u=1} \frac{\bar{\sigma}_u^2}{\varepsilon^2}\left[\frac{1}{|\mathcal{N}_{u,1}|}+\frac{1}{|\mathcal{N}_{u,2}|}\right]\label{eqn M prob PCSM 1}\\
&= \frac{\sum_u \bar{\sigma}_u^2}{U}\frac{U}{\varepsilon^2}\frac{2}{\Theta(N)}\label{eqn M prob PCSM 2}\\
&=  \frac{\sum_u \bar{\sigma}_u^2}{U}\frac{U}{\omega(U/N)}\frac{2}{\Theta(N)},\label{eqn M prob PCSM 3}
\end{align}
\end{subequations}
where \eqref{eqn M prob PCSM 1} follows by \eqref{eqn chebyshev PCSM}, and \eqref{eqn M prob PCSM 2} follows as $|\mathcal{N}_{u,1}|=|\mathcal{N}_{u,2}|=\Theta(N)$. Because \eqref{eqn M prob PCSM 3} approaches 0 for $U$ and $N$ both sufficiently large, Lemma \ref{lemma achievable bound M PCSM} can be verified.

\bibliographystyle{IEEEtran}     
\bibliography{IEEEabrv,strings}

\begin{IEEEbiographynophoto}{Fan Xu} received the B.S. degree in physics and the Ph.D. degree in information and communication engineering from Shanghai Jiao Tong University, Shanghai,
China, in 2016 and 2022, respectively. He received Huawei Scholarship in 2018 and was the outstanding graduate of Shanghai Jiao Tong University in 2022. 

Since 2022, he joined Peng Cheng Laboratory, Shenzhen, China, as a post-doctor. His research interests include coded caching, distributed computing, intelligent reflecting surface, signal processing and optimization of 5G and beyond networks.
\end{IEEEbiographynophoto}

\begin{IEEEbiographynophoto}{Jiawei Yao} received the M.Phil. degree and the B.Eng. degree in Computer and Information Engineering and Telecommunication Engineering from The Chinese University of Hong Kong, Shenzhen and Zhengzhou University in 2024 and 2021, respectively. His research interests include intelligent reflecting surface and integrated sensing and communication.
\end{IEEEbiographynophoto}

\begin{IEEEbiographynophoto}{Wenhai Lai} received the B.E. degree in information engineering from Beijing University of Posts and Telecommunications, in 2021. He is currently working toward the Ph.D. degree with the School of Science and Engineering, The Chinese University of Hong Kong, Shenzhen, China. His research interests include intelligent reflecting surface and reinforcement learning.
\end{IEEEbiographynophoto}

\begin{IEEEbiographynophoto}{Kaiming Shen} received the B.Eng. degree in information security and the B.Sc. degree in mathematics from Shanghai Jiao Tong University, China in 2011, and then the M.A.Sc. degree in electrical and computer engineering from the University of Toronto, Canada in 2013. After working at a tech startup in Ottawa for one year, he returned to the University of Toronto and received the Ph.D. degree in electrical and computer engineering in early 2020. Dr. Shen has been with the School of Science and Engineering at The Chinese University of Hong Kong (CUHK), Shenzhen, China as a tenure-track assistant professor since 2020. His research interests include optimization, wireless communications, information theory, and machine learning.

Dr. Shen received the IEEE Signal Processing Society Young Author Best Paper Award in 2021, the CUHK Teaching Achievement Award in 2023, and the Frontiers of Science Award at the International Congress of Basic Science in 2024. Dr. Shen currently serves as an Editor for IEEE Transactions on Wireless Communications.
\end{IEEEbiographynophoto}

\begin{IEEEbiographynophoto}{Xin Li}
graduated from Xidian University and joined Huawei in 2008. He has rich experience in wireless channel modeling and wireless network performance modeling and optimization. Currently, he is a technical expert in Huawei's experience lab, focusing on future-oriented network technology research, including new technologies such as Intelligent Reflection Surface and Intelligent Transmission Surface, and their application in network structure optimization.
\end{IEEEbiographynophoto}

\begin{IEEEbiographynophoto}{Xin Chen}
graduated from the Radio Engineering Department of Southeast University and joined Huawei in 2000. He has 20 years of R\&D experience in the wireless communications field. He has served as senior algorithm engineer, solution architect, and technology development director successively. He has rich experience and achievements in network planning, optimization, and operation and maintenance of mobile communication networks.

Currently, he is the director of the Algorithm Dept of the Service and Software R\&D domain of Huawei Carrier BG. He is responsible for the research of key technologies for digitalization and intelligence in the telecom field, including intelligent network optimization, autonomous network architecture and O\&M, analysis algorithms of telecom big data, and next-generation computing architecture in the telecom field.
\end{IEEEbiographynophoto}

\begin{IEEEbiographynophoto}{Zhi-Quan (Tom) Luo} is the Vice President (Academic) of The Chinese University of Hong Kong, Shenzhen where he has been a professor since 2014. He is concurrently the Director of Shenzhen Research Institute of Big Data. 

Professor Luo received his Ph.D. in Operations Research from MIT in 1989 and his B.S. degree in Mathematics in 1984 from Peking University, China. His research interests lie in the area of optimization, big data, signal processing and digital communication, ranging from theory, algorithms to design and implementation. He served as the Chair of the IEEE Signal Processing Society Technical Committee on Signal Processing for Communications (SPCOM) and the Editor in Chief for IEEE Transactions on Signal Processing (2012--2014), and was an Associate Editor for many internationally recognized journals.  

Professor Luo is a Fellow of the Institute of Electrical and Electronics Engineers (IEEE) and the Society for Industrial and Applied Mathematics (SIAM). He received the 2010 Farkas Prize from the INFORMS Optimization Society, and the 2018 Paul Y. Tseng Memorial Lectureship from the Mathematical Optimization Society. He also received three Best Paper Awards in 2004, 2009 and 2011, a Best Magazine Paper Award in 2015, all from the IEEE Signal Processing Society, and a 2011 Best Paper Award from the EURASIP. In 2014, he was elected to the Royal Society of Canada. Professor Luo was elected to the Chinese Academy of Engineering (as a foreign member) in 2021, and was awarded the Wang Xuan Applied Mathematics Prize in 2022 by the China Society of Industrial and Applied Mathematics.
\end{IEEEbiographynophoto}

\end{document}